\documentclass{article}
\usepackage{amsfonts}
\usepackage{amssymb}
\usepackage{graphicx}
\usepackage{amsmath}

\begin{document}

\title{On Shannon entropies in ${\mu}$-deformed Segal-Bargmann analysis}

\author{Claudio de Jes\'{u}s Pita Ruiz Velasco \thanks{Research partially supported
by
grant 165628, CONACyT (Mexico)}\\Universidad Panamericana\\Mexico City,
Mexico\\email: cpita@mx.up.mx \setcounter{footnote}{6}
\and Stephen Bruce Sontz \thanks{Research partially supported by grants 32146-E and
P-42227-F, CONACyT (Mexico)}\\Centro de Investigaci\'{o}n en Matem\'{a}ticas, A.C.
(CIMAT)\\Guanajuato, Mexico\\email: sontz@cimat.mx}
\date{October, 2005}

\maketitle

\begin{abstract}

\noindent We consider a ${\mu}$-deformation of the Segal-Bargmann transform,
which is a unitary map from a ${\mu}$-deformed quantum configuration space
onto a ${\mu}$-deformed quantum phase space (the ${\mu}$-deformed
Segal-Bargmann space). Both of these Hilbert spaces have canonical orthonormal
bases. We obtain explicit formulas for the Shannon entropy of some of the
elements of these bases. We also consider two reverse log-Sobolev inequalities
in the ${\mu}$-deformed Segal-Bargmann space, which have been proved in a
previous work, and show that a certain known coefficient in them is the best
possible.

\end{abstract}

\tableofcontents

\section{Introduction}

The Segal-Bargmann space $\mathcal{B}^{2}$ is the holomorphic subspace of the
Hilbert space $L^{2}\left(  \mathbb{C},d\nu_{\text{Gauss}}\right)  $, where
$d\nu_{\text{Gauss}}$ is a Gaussian measure. Since $\mathcal{B}^{2}$ is closed
in $L^{2}\left(  \mathbb{C},d\nu_{\text{Gauss}}\right)  $, the Segal-Bargmann
space is itself a Hilbert space. It is common to think of the Segal-Bargmann
space as a quantum phase space, similarly as one thinks of the space
$L^{2}\left(  \mathbb{R},dx\right)  $ as a quantum configuration space. The so
called Bargmann transform $\widetilde{B}:L^{2}\left(  \mathbb{R},dx\right)
\rightarrow\mathcal{B}^{2}$ is an isomorphism between these two quantum spaces
and \textit{Segal-Bargmann analysis} has to do mainly with the study of
operators related to $\widetilde{B}$ and spaces of holomorphic functions
related to $\mathcal{B}^{2}$. (The beginnings of this mathematical theory date
back to the works of Segal [Seg1], [Seg2] and Bargmann [Bar]. The physical
theory begins with the work of Fock [F].) The quantum configuration space can
be replaced by another unitarily equivalent space, namely $L^{2}%
(\mathbb{R},dg)$, called the \textit{ground state representation}, where $dg$
is another Gaussian measure. In this case, the resulting transform $B$ that
maps the ground state representation unitarily onto the Segal-Bargmann space
is called the \textit{Segal-Bargmann transform}. In both quantum spaces
$L^{2}(\mathbb{R},dg)$ and $\mathcal{B}^{2}$ there are defined unbounded
self-adjoint operators $Q$ (position) and $P$ (momentum), which satisfy the
relation $[P,Q]=-iI$, called the \textit{canonical commutation relation}
(CCR). The CCR implies the \textit{equations of motion} $i[P,H]=Q$ and
$i[Q,H]=-P$, where $H=2^{-1}(Q^{2}+P^{2})$ is the Hamiltonian of the harmonic
oscillator. In 1950, Wigner [Wig] proved that the converse implication is
false by exhibiting a family of unbounded operators, labeled by a parameter
$\mu>-1/2$, that satisfy the equations of motion but do not satisfy the CCR.
Rosenblum and Marron described explicitly (in [Ros1], [Ros2] and [Marr]) a
$\mu$-quantum configuration space $L^{2}(\mathbb{R},\left\vert x\right\vert
^{2\mu}dx)$, a $\mu$-Segal-Bargmann space $\mathcal{B}_{\mu}^{2}$, and a $\mu
$-Bargmann transform $\widetilde{B}_{\mu}$ which is a unitary onto
transformation mapping the former Hilbert space to the latter Hilbert space.
This theory can be understood as a $\mu$-deformation of standard
Segal-Bargmann analysis with the property that if one sets $\mu=0$ the
standard theory is recovered (see [Snt3]). So we will refer to $L^{2}%
(\mathbb{R},\left\vert x\right\vert ^{2\mu}dx)$ and $\mathcal{B}_{\mu}^{2}$,
as the \textquotedblleft$\mu$-deformed quantum configuration
space\textquotedblright\ and the \textquotedblleft$\mu$-deformed
Segal-Bargmann space\textquotedblright, respectively, and to $\widetilde
{B}_{\mu}$ as the \textquotedblleft$\mu$-deformed Bargmann
transform\textquotedblright. It is easy to obtain explicitly also the
\textquotedblleft$\mu$-deformed ground state representation\textquotedblright%
\ $L^{2}(\mathbb{R},dg_{\mu})$ and the \textquotedblleft$\mu$-deformed
Segal-Bargmann transform\textquotedblright\ $B_{\mu}$, which is a unitary map
from $L^{2}(\mathbb{R},dg_{\mu})$ onto $\mathcal{B}_{\mu}^{2}$.

In his paper [Snt1] the second author put emphasis on the Shannon entropy (to
be defined in Section 2) as an important quantity in Segal-Bargmann analysis.
More precisely, following [Hir] the second author proved a log-Sobolev
inequality, where the entropies of a function $f\in L^{2}(\mathbb{R},dg)$ and
of its Segal-Bargmann transform $Bf\in\mathcal{B}^{2}$ are involved. Later in
[Snt2], the second author obtained explicit formulas for the entropy of
relevant elements of the Hilbert spaces $L^{2}(\mathbb{R},dg)$ and
$\mathcal{B}^{2}$, namely, elements of the corresponding canonical basis of
these spaces. By denoting by $\zeta_{n}$, $n=0,1...$ the functions of the
canonical basis $\left\{  \zeta_{n}\right\}  _{n=0}^{\infty}$ of the ground
state representation $L^{2}(\mathbb{R},dg)$, and by $\xi_{n}$, $n=0,1...$ the
functions of the canonical basis $\left\{  \xi_{n}\right\}  _{n=0}^{\infty}$
of Segal-Bargmann space $\mathcal{B}^{2}$, the second author proved in [Snt2] that%

\begin{equation}
S_{L^{2}\left(  \mathbb{C},d\nu_{\text{Gauss}}\right)  }\left(  \xi
_{n}\right)  =n\left(  -\gamma+1+\frac{1}{2}+\cdots+\frac{1}{n}\right)  -\log
n!,\tag{1.1}%
\end{equation}%
\begin{equation}
S_{L^{2}\left(  \mathbb{R},dg\right)  }\left(  \zeta_{1}\right)
=2-\log2-\gamma,\tag{1.2}%
\end{equation}
where $S_{L^{2}\left(  \mathbb{C},d\nu_{\text{Gauss}}\right)  }\left(  \xi
_{n}\right)  $ is the entropy of $\xi_{n}\in\mathcal{B}^{2}$, $n=0,1,...$,
$S_{L^{2}\left(  \mathbb{R},dg\right)  }\left(  \zeta_{1}\right)  $ is the
entropy of $\zeta_{1}\in L^{2}(\mathbb{R},dg)$, and $\gamma$ is Euler's constant.

In the context of the $\mu$-deformed theory of Segal-Bargmann analysis,
similar results to those in [Snt1] have been recently proven, e.g. log-Sobolev
and reverse log-Sobolev inequalities. (See [A-S.1], [A-S.2] and [P-S].) What
we want to do in this work is to obtain, for the $\mu$-deformed theory,
similar results to those in [Snt2]. That is, we want to obtain explicit
formulas for the entropies of the $\mu$-deformed elements $\zeta_{1}^{\mu} $
and $\xi_{n}^{\mu}$, $n=0,1,...$ of the corresponding $\mu$-deformed canonical
basis $\left\{  \zeta_{n}^{\mu}\right\}  _{n=0}^{\infty}$ and $\left\{
\zeta_{n}^{\mu}\right\}  _{n=0}^{\infty}$ of the $\mu$-deformed Hilbert spaces
$L^{2}(\mathbb{R},dg_{\mu})$ and $\mathcal{B}_{\mu}^{2}$, respectively.

We now outline the content of the work. In Section 2 we give the definitions
and notation that will be used throughout the work. In this section we also
introduce the $\mu$-deformed Hilbert spaces $L^{2}(\mathbb{R},dg_{\mu})$ and
$\mathcal{B}_{\mu}^{2}$, and their canonical bases as well. In Section 3 we
give some preliminary results that will help us to analyze some properties of
the sequence of entropies of the functions $\xi_{n}^{\mu}$, $n=0,1,...$. These
properties are not explicitly given (in the case $\mu=0$) in [Snt2], but we
give them as a proposition at the end of Section 3. In Section 4 we obtain
explicit formulas for the entropies of the elements $\xi_{n}^{\mu}%
\in\mathcal{B}_{\mu}^{2}$, $n=1,2,...$, and we study some properties of the
corresponding sequence of entropies. The results in this section generalize
the formula (1.1) of [Snt2], as well as the proposition at the end of Section
3 mentioned above. In Section 5 we consider the $\mu$-deformed ground state
representation and we obtain explicit formulas for the monomials $t^{n}\in
L^{2}(\mathbb{R},dg_{\mu})$, $n=0,1,...$. Unfortunately the technique we use
here to obtain these formulas (and those of Section 4) does not work to obtain
the entropies of the elements $\zeta_{n}^{\mu}$ of the canonical basis of
$L^{2}(\mathbb{R},dg_{\mu})$, for $n\geq2$. It turns out that our method for
calculating the entropy of a function $f$ works only in the case of $f$ being
a monomial, and the elements $\zeta_{n}^{\mu} $ are monomials only for
$n=0,1$. The formula we obtain for the entropy of $\zeta_{1}^{\mu}$
generalizes the formula (1.2) of [Snt2].

Also, by means of a concrete example, in Section 5 we show that the $\mu
$-deformed Segal-Bargmann transform $B_{\mu}$ does not preserve entropy. In
Section 6 we consider two reverse log-Sobolev inequalities proved in [A-S.2],
in which the condition $c>1$ of a certain parameter $c$ appears as a
sufficient condition. In this section we show that this condition is also
necessary, or in other words, that the condition $c>1$ is the best possible.
Finally, in Section 7 we make some comments about what we left unfinished in
this paper and what is possible to do beyond the results presented here.

\bigskip

\section{Definitions and notation}

\bigskip

In this section we give the definitions and the notation that we will use
throughout the work. First, we take $\mu>-\frac{1}{2}$ to be a fixed parameter
(unless otherwise stated). The (Coxeter) group $\mathbb{Z}_{2}$ is the
multiplicative group $\left\{  -1,1\right\}  $, and $\log$ is the natural
logarithm (base $e$). We use the convention $0\log0=0$ (which makes the
function $\phi:\left[  0,\infty\right)  \rightarrow\mathbb{R}$, $\phi\left(
x\right)  =x\log x$ continuous). We also use the convention that $C$ denotes a
constant (a quantity that does not depend on the variables of interest in the
context), which may change its value every time it appears. We denote by
$\mathcal{H}\left(  \mathbb{C}\right)  $ the space of holomorphic functions
$f:\mathbb{C\rightarrow C}$ with the topology of uniform convergence on
compact sets.

We begin by defining the $\mu$-deformations of the factorial function and of
the exponential function. Let $\mathbb{N}$ denote the set of positive integers.

\bigskip

\textbf{Definition 2.1 }\textit{The }$\mu$\textit{-deformed factorial function
}$\gamma_{\mu}:\mathbb{N}\cup\left\{  0\right\}  \rightarrow\mathbb{R}%
$\textit{\ is defined by }$\gamma_{\mu}\left(  0\right)  =1$\textit{\ and }%
\[
\gamma_{\mu}\left(  n\right)  :=\left(  n+2\mu\theta\left(  n\right)  \right)
\gamma_{\mu}\left(  n-1\right)  ,
\]
\textit{\ where }$n\in\mathbb{N}$\textit{\ and }$\theta:\mathbb{N}%
\rightarrow\left\{  0,1\right\}  $\textit{\ is the characteristic function of
the odd positive integers. The }$\mu$\textit{-deformed} \textit{exponential
function }$\mathbf{e}_{\mu}:\mathbb{C}\rightarrow\mathbb{C}$\textit{, is
defined by the power series }%
\[
\mathbf{e}_{\mu}\left(  z\right)  :=\sum_{n=0}^{\infty}\frac{z^{n}}%
{\gamma_{\mu}\left(  n\right)  }.
\]

\bigskip

We note that $\gamma_{0}\left(  n\right)  =n!$ (the usual factorial function)
and so $\mathbf{e}_{0}\left(  z\right)  =\exp\left(  z\right)  $ (the usual
complex exponential function). It is clear that the power series in the
definition of $\mathbf{e}_{\mu}\left(  z\right)  $ is absolutely convergent
for all $z\in\mathbb{C}$. So the $\mu$-deformed exponential $\mathbf{e}_{\mu}$
is an entire function.

We will use the following explicit formulas for $\gamma_{\mu}\left(
2n\right)  $ and $\gamma_{\mu}\left(  2n+1\right)  $, $n=0,1,2,...$ (see
[Ros1], p. 371):
\begin{align}
\gamma_{\mu}(2n)  & =\frac{2^{2n}\Gamma\left(  n+1\right)  \Gamma\left(
\mu+n+\frac{1}{2}\right)  }{\Gamma\left(  \mu+\frac{1}{2}\right)  }\tag{2.1}\\
& =\frac{\left(  2n\right)  !\Gamma\left(  \frac{1}{2}\right)  \Gamma\left(
\mu+n+\frac{1}{2}\right)  }{\Gamma\left(  \mu+\frac{1}{2}\right)
\Gamma\left(  n+\frac{1}{2}\right)  },\nonumber
\end{align}%
\begin{align}
\gamma_{\mu}(2n+1)  & =\frac{2^{2n+1}\Gamma\left(  n+1\right)  \Gamma\left(
\mu+n+\frac{3}{2}\right)  }{\Gamma\left(  \mu+\frac{1}{2}\right)  }\tag{2.2}\\
& =\frac{\left(  2n+1\right)  !\Gamma\left(  \frac{1}{2}\right)  \Gamma\left(
\mu+n+\frac{3}{2}\right)  }{\Gamma\left(  \mu+\frac{1}{2}\right)
\Gamma\left(  n+\frac{3}{2}\right)  }.\nonumber
\end{align}

\bigskip

The following definition (from [Ros1]) gives us a $\mu$-deformation of the
classical Hermite polynomials.

\bigskip

\textbf{Definition 2.2 \ }\textit{For }$n=0,1,...$\textit{\ we define the }$n
$\textit{-th }$\mu$\textit{-deformed Hermite polynomial} $H_{n}^{\mu}%
(t)$\textit{\ by the generating function}%
\[
\exp\left(  -z^{2}\right)  \mathbf{e}_{\mu}\left(  2tz\right)  =\sum
_{n=0}^{\infty}H_{n}^{\mu}\left(  t\right)  \frac{z^{n}}{n!}.
\]

\bigskip

It is easy to check that $H_{n}^{\mu}\left(  t\right)  $ is in fact a
polynomial of degree $n$ in the real variable $t$. For example, we have that
$H_{0}^{\mu}\left(  t\right)  =1$, $H_{1}^{\mu}\left(  t\right)  =\frac
{2}{1+2\mu}t$, $H_{2}^{\mu}\left(  t\right)  =\frac{4}{1+2\mu}t^{2}-2$, and so on.

The normalized $\mu$-deformed Hermite polynomials $\zeta_{n}^{\mu}\left(
t\right)  $, $n=0,1,...$ defined by%
\begin{equation}
\zeta_{n}^{\mu}\left(  t\right)  :=2^{-\frac{n}{2}}\left(  n!\right)
^{-1}\left(  \gamma_{\mu}(n)\right)  ^{\frac{1}{2}}H_{n}^{\mu}\left(
t\right)  ,\tag{2.3}%
\end{equation}
form an orthonormal basis of the $\mu$-deformed ground state representation
\linebreak$L^{2}\left(  \mathbb{R},dg_{\mu}\right)  $, where $dg_{\mu}$ is the
$\mu$\textit{-deformed Gaussian measure }defined by%
\begin{equation}
dg_{\mu}\left(  t\right)  :=\left(  \Gamma\left(  \mu+\frac{1}{2}\right)
\right)  ^{-1}\exp\left(  -t^{2}\right)  \left\vert t\right\vert ^{2\mu
}dt.\tag{2.4}%
\end{equation}

The basis $\left\{  \zeta_{n}^{\mu}\right\}  _{n=0}^{\infty}$ is called
\textit{the canonical basis of }$L^{2}\left(  \mathbb{R},dg_{\mu}\right)  $.
(See [Ros1] and [P-S].)

The case $\mu=0$ recovers the well known fact that for $n=0,1,...$, the
normalized polynomials $\zeta_{n}\left(  t\right)  =2^{-\frac{n}{2}}\left(
n!\right)  ^{-\frac{1}{2}}H_{n}\left(  t\right)  $, where $H_{n}\left(
t\right)  $ denotes the \linebreak$n$-th Hermite polynomial, form the
canonical orthonormal basis of the ground state representation $L^{2}\left(
\mathbb{R},dg\right)  $, where $dg$ is the Gaussian probability measure
$dg\left(  t\right)  =\pi^{-\frac{1}{2}}\exp\left(  -t^{2}\right)  dt$. (See [Hall].)

\bigskip

\textbf{Definition 2.3 \ }\textit{We define the measure }$d\nu_{\mu}%
$\textit{\ on the space }$\mathbb{C}\times\mathbb{Z}_{2}$ \textit{by}
\begin{equation}
d\nu_{\mu}\left(  z,1\right)  :=\frac{2^{\frac{1}{2}-\mu}}{\pi\Gamma\left(
\mu+\frac{1}{2}\right)  }K_{\mu-\frac{1}{2}}\left(  \left\vert z\right\vert
^{2}\right)  \left\vert z\right\vert ^{2\mu+1}dxdy,\tag{2.5}%
\end{equation}%
\begin{equation}
d\nu_{\mu}\left(  z,-1\right)  :=\frac{2^{\frac{1}{2}-\mu}}{\pi\Gamma\left(
\mu+\frac{1}{2}\right)  }K_{\mu+\frac{1}{2}}\left(  \left\vert z\right\vert
^{2}\right)  \left\vert z\right\vert ^{2\mu+1}dxdy,\tag{2.6}%
\end{equation}
\textit{where }$\Gamma$\textit{\ is the Euler gamma function, }$K_{\alpha}%
$\textit{\ is the Macdonald function of order }$\alpha$\textit{\ (both defined
in }[Leb]\textit{), and }$dxdy$\textit{\ is Lebesgue measure on }$\mathbb{C}%
$\textit{.}

\bigskip

By using that $\mathbb{C\cong C\times}\left\{  1\right\}  \cong\mathbb{C\times
}\left\{  -1\right\}  $, we will identify the restrictions (2.5) and (2.6) as
measures on $\mathbb{C}$.

The Macdonald function $K_{\alpha}$ is the modified Bessel function of the
third kind (with purely imaginary argument, as described in [Wat], p. 78),
which is known to be a holomorphic function on $\mathbb{C}\setminus\left(
-\infty,0\right]  $ and is entire with respect to the parameter $\alpha$.
Nevertheless, our interest will be only in the values and behavior of this
function for $x\in\mathbb{R}^{+}$ and $\alpha\in\mathbb{R}$. For
$z\in\mathbb{C}$, $\left\vert \arg z\right\vert <\pi$ and $\alpha
\notin\mathbb{Z}$, the Macdonald function can be defined as
\[
K_{\alpha}\left(  z\right)  =\frac{\pi}{2}\frac{I_{-\alpha}\left(  z\right)
-I_{\alpha}\left(  z\right)  }{\sin\left(  \alpha\pi\right)  }%
\]
(see [Leb], p. 108), where $I_{\alpha}\left(  z\right)  $ is the modified
Bessel function of the first kind. For $\alpha\in\mathbb{Z}$, we define
$K_{\alpha}\left(  z\right)  =\lim_{\beta\rightarrow\alpha}K_{\beta}\left(
z\right)  $. This expression shows that $K_{a}\left(  z\right)  $ is an even
function of the parameter $\alpha$. In particular, since $I_{\frac{1}{2}%
}\left(  z\right)  =\left(  \frac{2}{\pi z}\right)  ^{\frac{1}{2}}\sinh z$ and
$I_{-\frac{1}{2}}\left(  z\right)  =\left(  \frac{2}{\pi z}\right)  ^{\frac
{1}{2}}\cosh z$ (see [Leb], p. 112), we have that
\[
K_{\pm\frac{1}{2}}\left(  z\right)  =\left(  \frac{\pi}{2z}\right)  ^{\frac
{1}{2}}\exp\left(  -z\right)  \text{,}%
\]
which shows that for $\mu=0$ the measures defined on $\mathbb{C}$ by (2.5) and
(2.6) are the same Gaussian measure:
\[
d\nu_{0}\left(  z,1\right)  =d\nu_{0}\left(  z,-1\right)  =\pi^{-1}\exp\left(
-\left\vert z\right\vert ^{2}\right)  dxdy,
\]
which is the Gaussian measure $d\nu_{\text{Gauss}}$ of the Segal-Bargmann
space $\mathcal{B}^{2}=\mathcal{H}\left(  \mathbb{C}\right)  \cap L^{2}\left(
\mathbb{C},d\nu_{\text{Gauss}}\right)  $.

By using the formula%
\begin{equation}
\int_{0}^{\infty}K_{\alpha}\left(  s\right)  s^{\beta-1}ds=2^{\beta-2}%
\Gamma\left(  \frac{\beta-\alpha}{2}\right)  \Gamma\left(  \frac{\beta+\alpha
}{2}\right)  ,\tag{2.7}%
\end{equation}
which holds if $\operatorname{Re}\beta>\left\vert \operatorname{Re}%
\alpha\right\vert $ (see [Wat], p. 388), we can see that (2.5) and (2.6) are
finite measures on $\mathbb{C}$, and moreover that the former is a probability
measure. (See [P-S].)

The integral representation%
\begin{equation}
K_{\alpha}\left(  z\right)  =\int_{0}^{\infty}\exp\left(  -z\cosh u\right)
\cosh\left(  \alpha u\right)  du\qquad\operatorname{Re}z>0\tag{2.8}%
\end{equation}
(see [Leb], p. 119) gives us at once two important properties of the Macdonald
function. The first is that $K_{\alpha}\left(  x\right)  >0$ for all
$x\in\mathbb{R}^{+}$, and the second is that $K_{\alpha}$ is a monotone
decreasing function for $x\in\mathbb{R}^{+}$.

We will work with the Hilbert space $L^{2}\left(  \mathbb{C}\times
\mathbb{Z}_{2},d\nu_{\mu}\right)  $. The norm of a vector $f\in L^{2}\left(
\mathbb{C}\times\mathbb{Z}_{2},d\nu_{\mu}\right)  $ will be denoted by
$\left\Vert f\right\Vert _{L^{2}\left(  \mathbb{C}\times\mathbb{Z}_{2}%
,d\nu_{\mu}\right)  }$. Let us consider the space
\[
\mathfrak{H}_{2,\mu}\!\!=\!\!\left\{  f:\mathbb{C}\rightarrow\mathbb{C}\mid
f_{e}\in L^{2}(\mathbb{C},\left.  d\nu_{\mu}\right\vert _{\mathbb{C}%
\times\left\{  1\right\}  })\text{ and }f_{o}\in L^{2}(\mathbb{C},\left.
d\nu_{\mu}\right\vert _{\mathbb{C}\times\left\{  -1\right\}  })\right\}  ,
\]
where $f=f_{e}+f_{o}$ is the decomposition of $f$ into its even and odd parts.
Observe that when $\mu=0$ we have $\mathfrak{H}_{2,0}\!\!=L^{2}\left(
\mathbb{C},d\nu_{\text{Gauss}}\right)  $.

For $f\in\mathfrak{H}_{2,\mu}$ we define
\[
\left\Vert f\right\Vert _{\mathfrak{H}_{2,\mu}}^{2}:=\left\Vert f_{e}%
\right\Vert _{L^{2}\left(  \mathbb{C},\left.  d\nu_{\mu}\right\vert
_{\mathbb{C}\times\left\{  1\right\}  }\right)  }^{2}+\left\Vert
f_{o}\right\Vert _{L^{2}\left(  \mathbb{C},\left.  d\nu_{\mu}\right\vert
_{\mathbb{C}\times\left\{  -1\right\}  }\right)  }^{2}.
\]

The linear map $\Phi:\mathfrak{H}_{2,\mu}\rightarrow L^{2}\left(
\mathbb{C}\times\mathbb{Z}_{2},d\nu_{\mu}\right)  $ defined as $\left(  \Phi
f\right)  \left(  z,1\right)  =f_{e}\left(  z\right)  $ and $\left(  \Phi
f\right)  \left(  z,-1\right)  =f_{o}\left(  z\right)  $ is injective and has
the property that
\begin{equation}
\left\Vert f\right\Vert _{\mathfrak{H}_{2,\mu}}=\left\Vert \Phi f\right\Vert
_{L^{2}\left(  \mathbb{C}\times\mathbb{Z}_{2},d\nu_{\mu}\right)  }\tag{2.9}%
\end{equation}
for all $f\in\mathfrak{H}_{2,\mu}$. Therefore $\left\Vert \cdot\right\Vert
_{\mathfrak{H}_{2,\mu}}$ is a norm on $\mathfrak{H}_{2,\mu}$. It is not hard
to show that the range of $\Phi$ is a closed subspace of $L^{2}\left(
\mathbb{C}\times\mathbb{Z}_{2},d\nu_{\mu}\right)  $. Therefore $\mathfrak{H}%
_{2,\mu}$ is a Hilbert space, since we have identified it with a closed
subspace of the Hilbert space $L^{2}\left(  \mathbb{C}\times\mathbb{Z}%
_{2},d\nu_{\mu}\right)  $. For a function $f\in\mathfrak{H}_{2,\mu}$ we will
sometimes write its norm as $\left\Vert f\right\Vert _{L^{2}\left(
\mathbb{C}\times\mathbb{Z}_{2},d\nu_{\mu}\right)  }$, meaning that we are
using (2.9) and identifying $f$ with $\Phi f$.

We will use the notations $d\nu_{e,\mu}$ and $d\nu_{o,\mu}$ for the
restrictions $\left.  d\nu_{\mu}\right\vert _{\mathbb{C}\times\left\{
1\right\}  }$ and $\left.  d\nu_{\mu}\right\vert _{\mathbb{C}\times\left\{
-1\right\}  }$, respectively. So for $f\in\mathfrak{H}_{2,\mu}$ we have
\begin{align*}
\left\Vert f\right\Vert _{\mathfrak{H}_{2,\mu}}^{2}  & =\left\Vert
f_{e}\right\Vert _{L^{2}\left(  \mathbb{C},d\nu_{e,\mu}\right)  }%
^{2}+\left\Vert f_{o}\right\Vert _{L^{2}\left(  \mathbb{C},d\nu_{o,\mu
}\right)  }^{2}\\
& =\left\Vert f_{e}\right\Vert _{\mathfrak{H}_{2,\mu}}^{2}+\left\Vert
f_{o}\right\Vert _{\mathfrak{H}_{2,\mu}}^{2}.
\end{align*}

\bigskip

\textbf{Definition 2.4 }\textit{The }$\mu$\textit{-deformed Segal-Bargmann
space, denoted by }$\mathcal{B}_{\mu}^{2}$\textit{, is defined as}
\begin{equation}
\mathcal{B}_{\mu}^{2}:=\mathcal{H}\left(  \mathbb{C}\right)  \cap
\mathfrak{H}_{2,\mu}.\tag{2.10}%
\end{equation}

\bigskip

That is, $\mathcal{B}_{\mu}^{2}$ is the holomorphic subspace of $\mathfrak{H}%
_{2,\mu}$. It turns out that $\mathcal{B}_{\mu}^{2}$ is closed in
$\mathfrak{H}_{2,\mu}$, and then it is also closed in $L^{2}\left(
\mathbb{C}\times\mathbb{Z}_{2},d\nu_{\mu}\right)  $, so $\mathcal{B}_{\mu}%
^{2}$ is itself a Hilbert space. (The proof of this fact does not depend on
$\mu$; see Theorem 2.2 in [Hall] for the case $\mu=0$.) Observe that when
$\mu=0$ we have $\mathcal{B}_{0}^{2}=\mathcal{H}\left(  \mathbb{C}\right)
\cap\mathfrak{H}_{2,0}=\mathcal{H}\left(  \mathbb{C}\right)  \cap L^{2}\left(
\mathbb{C},d\nu_{\text{Gauss}}\right)  =\mathcal{B}^{2}$.

If we decompose the space $\mathcal{H}\left(  \mathbb{C}\right)  $ of
holomorphic functions $f:\mathbb{C\rightarrow C}$ as $\mathcal{H}\left(
\mathbb{C}\right)  =\mathcal{H}_{e}\left(  \mathbb{C}\right)  \oplus
\mathcal{H}_{o}\left(  \mathbb{C}\right)  $, where
\begin{align*}
\mathcal{H}_{e}\left(  \mathbb{C}\right)  \!\!:=  & \left\{  f\in
\mathcal{H}\left(  \mathbb{C}\right)  :f=f_{e}\right\} \\
\text{and\quad}\ \mathcal{H}_{o}\left(  \mathbb{C}\right)  \!\!:=  & \left\{
f\in\mathcal{H}\left(  \mathbb{C}\right)  :f=f_{o}\right\}
\end{align*}
are the subspaces of the even and odd functions of $\mathcal{H}\left(
\mathbb{C}\right)  $, respectively, then by writing $\mathcal{H}\left(
\mathbb{C}\right)  \ni f=f_{e}+f_{o}$, the space $\mathcal{B}_{\mu}^{2}$ is
just the space of holomorphic functions $f:\mathbb{C\rightarrow C}$ such that
the even part $f_{e}$ (the odd part $f_{o}$) of $f$ is square integrable with
respect to the measure $d\nu_{e,\mu}$ (with respect to the measure
$d\nu_{o,\mu}$, respectively). That is,
\[
\mathcal{B}_{\mu}^{2}=\left\{  f\in\mathcal{H}\left(  \mathbb{C}\right)
:f_{e}\in L^{2}\left(  \mathbb{C},d\nu_{e,\mu}\right)  \text{ and }f_{o}\in
L^{2}\left(  \mathbb{C},d\nu_{o,\mu}\right)  \right\}  .
\]

Yet another way to think of $\mathcal{B}_{\mu}^{2}$ is as
\begin{equation}
\mathcal{B}_{\mu}^{2}=\mathcal{B}_{e,\mu}^{2}\oplus\mathcal{B}_{o,\mu}%
^{2},\tag{2.11}%
\end{equation}
where
\begin{align*}
\mathcal{B}_{e,\mu}^{2}\!\!  & =\!\!\mathcal{H}_{e}\left(  \mathbb{C}\right)
\cap\mathfrak{H}_{2,\mu}\\
\text{and}\ \quad\mathcal{B}_{o,\mu}^{2}\!\!  & =\mathcal{H}_{o}\left(
\mathbb{C}\right)  \cap\mathfrak{H}_{2,\mu}%
\end{align*}
are the even and odd subspaces of $\mathcal{B}_{\mu}^{2}$.

Observe that the inner product of the Hilbert space $\mathcal{B}_{\mu}^{2}$
(from which the norm on $\mathcal{B}_{\mu}^{2}$ defined above comes) is
\begin{equation}
\left\langle f,g\right\rangle _{\mathcal{B}_{\mu}^{2}}=\left\langle
f_{e},g_{e}\right\rangle _{L^{2}\left(  \mathbb{C},d\nu_{e,\mu}\right)
}+\left\langle f_{o},g_{o}\right\rangle _{L^{2}\left(  \mathbb{C},d\nu_{o,\mu
}\right)  }.\tag{2.12}%
\end{equation}

We then have that $\mathcal{B}_{e,\mu}^{2}$ and $\mathcal{B}_{o,\mu}^{2}$ are
orthogonal subspaces of $\mathcal{B}_{\mu}^{2}$, and that (2.11) holds as
Hilbert spaces.

The monomials $\xi_{n}^{\mu}\left(  z\right)  $, $n=0,1,...$ defined for $z
\in\mathbb{C}$ by%
\begin{equation}
\xi_{n}^{\mu}\left(  z\right)  :=\left(  \gamma_{\mu}(n)\right)  ^{-\frac
{1}{2}}z^{n},\tag{2.13}%
\end{equation}
form an orthonormal basis of the $\mu$-deformed Segal-Bargmann space
$\mathcal{B}_{\mu}^{2}$. The basis $\left\{  \xi_{n}^{\mu}\right\}
_{n=0}^{\infty}$ is called the \textit{canonical basis of }$\mathcal{B}_{\mu
}^{2}$. When $\mu=0$ we obtain the monomials $\xi_{n}\left(  z\right)
=\left(  n!\right)  ^{-\frac{1}{2}}z^{n}$, $n=0,1,...$ which are known to form
the canonical basis of the Segal-Bargmann space $\mathcal{B}^{2}$. (See [Hall].)

The $\mu$-deformed Segal-Bargmann transform $B_{\mu}:L^{2}\left(
\mathbb{R},dg_{\mu}\right)  \rightarrow\mathcal{B}_{\mu}^{2}$ can be defined
as $B_{\mu}\left(  \zeta_{n}^{\mu}\right)  =\xi_{n}^{\mu}$, $n=0,1,...$. It is
clear that $B_{\mu}$ so defined is a unitary map. An explicit formula for
$B_{\mu}$ is%
\begin{equation}
\left(  B_{\mu}f\right)  \left(  z\right)  =\exp\left(  -\frac{z^{2}}%
{2}\right)  \int_{\mathbb{R}}\mathbf{e}_{\mu}\left(  2^{\frac{1}{2}}tz\right)
f\left(  t\right)  dg_{\mu}\left(  t\right)  .\tag{2.14}%
\end{equation}
(See [P-S].) When $\mu=0$ this formula becomes
\[
\left(  B_{0}f\right)  (z)=\int_{\mathbb{R}}\exp\left(  -\frac{z^{2}}%
{2}+2^{\frac{1}{2}}tz\right)  f(t)dg\left(  t\right)  ,
\]
which is the undeformed\ Segal-Bargmann transform studied, for example, in
[Hall], where it is shown that it is a unitary map from the quantum
configuration space $L^{2}\left(  \mathbb{R},dg\right)  $ onto the quantum
phase space $\mathcal{B}^{2}$.

\bigskip

\textbf{Definition 2.5 \ }\textit{Let }$\left(  \Omega,d\nu\right)
$\textit{\ be a finite measure space, that is, $0<\nu(\Omega)<\infty$. For
}$f\in L^{2}\left(  \Omega,d\nu\right)  $\textit{, the Shannon entropy
}$S_{L^{2}\left(  \Omega,d\nu\right)  }\left(  f\right)  $\textit{\ is defined
by }
\begin{equation}
S_{L^{2}\left(  \Omega,d\nu\right)  }\left(  f\right)  :=\int_{\Omega
}\left\vert f\left(  \omega\right)  \right\vert ^{2}\log\left\vert f\left(
\omega\right)  \right\vert ^{2}d\nu\left(  \omega\right)  -\left\Vert
f\right\Vert _{L^{2}\left(  \Omega,d\nu\right)  }^{2}\log\left\Vert
f\right\Vert _{L^{2}\left(  \Omega,d\nu\right)  }^{2}.\tag{2.15}%
\end{equation}

\bigskip

This definition was introduced by Shannon [Sha] in his Theory of
Communication. Note that, since $\left(  \Omega,d\nu\right)  $ is a finite
measure space, the entropy $S_{L^{2}\left(  \Omega,d\nu\right)  }\left(
f\right)  $ makes sense for all $f\in L^{2}\left(  \Omega,d\nu\right)  $.
Moreover, by considering the convex function $\phi:\left[  0,\infty\right)
\rightarrow\mathbb{R}$, $\phi\left(  x\right)  =x\log x$, and the probability
measure space $\left(  \Omega,d\nu^{\prime}\right)  $, where $d\nu^{\prime
}=W^{-1}d\nu$, $W=\nu\left(  \Omega\right)  $, we have by Jensen's inequality
(see [L-L], p. 38) that
\[
\left(  \int_{\Omega}\left\vert f\left(  \omega\right)  \right\vert ^{2}%
d\nu\left(  \omega\right)  \right)  \!\log\!\left(  \frac{1}{W}\int_{\Omega
}\left\vert f\left(  \omega\right)  \right\vert ^{2}d\nu\left(  \omega\right)
\right)  \!\leq\!\int_{\Omega}\left\vert f\left(  \omega\right)  \right\vert
^{2}\log\left\vert f\left(  \omega\right)  \right\vert ^{2}d\nu\left(
\omega\right)
\]
or
\[
\left(  -\log W\right)  \left\Vert f\right\Vert _{L^{2}\left(  \Omega
,d\nu\right)  }^{2}\leq S_{L^{2}\left(  \Omega,d\nu\right)  }\left(  f\right)
,
\]
which shows that $S_{L^{2}\left(  \Omega,d\nu\right)  }\left(  f\right)
\neq-\infty$, though $S_{L^{2}\left(  \Omega,d\nu\right)  }\left(  f\right)
=+\infty$ can happen. Also observe that $S_{L^{2}\left(  \Omega,d\nu^{\prime
}\right)  }\left(  f\right)  \geq0$, though $S_{L^{2}\left(  \Omega
,d\nu\right)  }\left(  f\right)  $ can be negative. Finally, note that
$S_{L^{2}\left(  \Omega,d\nu\right)  }\left(  f\right)  $ is homogeneous of
degree $2$.

Observe that for $f\in\mathcal{B}_{\mu}^{2}$, $f\neq0$, the entropy
$S_{L^{2}\left(  \mathbb{C}\times\mathbb{Z}_{2},d\nu_{\mu}\right)  }\left(
f\right)  $ is\textit{\ not in general }equal to $S_{L^{2}\left(
\mathbb{C},d\nu_{e,\mu}\right)  }\left(  f_{e}\right)  +S_{L^{2}\left(
\mathbb{C},d\nu_{o,\mu}\right)  }\left(  f_{o}\right)  $. What we really have is%
\begin{align}
S_{L^{2}\left(  \mathbb{C}\times\mathbb{Z}_{2},d\nu_{\mu}\right)  }\left(
f\right)   & =S_{L^{2}\left(  \mathbb{C},d\nu_{e,\mu}\right)  }\left(
f_{e}\right)  +S_{L^{2}\left(  \mathbb{C},d\nu_{o,\mu}\right)  }\left(
f_{o}\right) \tag{2.16}\\
& +\left\Vert f_{e}\right\Vert _{L^{2}\left(  \mathbb{C},d\nu_{e,\mu}\right)
}^{2}\log\frac{\left\Vert f_{e}\right\Vert _{L^{2}\left(  \mathbb{C}%
,d\nu_{e,\mu}\right)  }^{2}}{\left\Vert f\right\Vert _{L^{2}\left(
\mathbb{C}\times\mathbb{Z}_{2},d\nu_{\mu}\right)  }^{2}}\nonumber\\
& +\left\Vert f_{o}\right\Vert _{L^{2}\left(  \mathbb{C},d\nu_{o,\mu}\right)
}^{2}\log\frac{\left\Vert f_{o}\right\Vert _{L^{2}\left(  \mathbb{C}%
,d\nu_{o,\mu}\right)  }^{2}}{\left\Vert f\right\Vert _{L^{2}\left(
\mathbb{C}\times\mathbb{Z}_{2},d\nu_{\mu}\right)  }^{2}}.\nonumber
\end{align}

Nevertheless, observe that if $f$ is an even (odd) function, its entropy is
given by $S_{L^{2}\left(  \mathbb{C},d\nu_{e,\mu}\right)  }\left(  f\right)  $
($S_{L^{2}\left(  \mathbb{C},d\nu_{o,\mu}\right)  }\left(  f\right)  $,
respectively). Then, for the functions $\xi_{n}^{\mu}$ of the canonical basis
of $\mathcal{B}_{\mu}^{2}$ we have $S_{n}^{\mu}=S_{L^{2}\left(  \mathbb{C}%
,d\nu_{e,\mu}\right)  }\left(  \xi_{n}^{\mu}\right)  $ if $n$ is even, and
$S_{n}^{\mu}=S_{L^{2}\left(  \mathbb{C},d\nu_{o,\mu}\right)  }\left(  \xi
_{n}^{\mu}\right)  $ if $n$ is odd, where $S_{n}^{\mu}:=S_{L^{2}\left(
\mathbb{C}\times\mathbb{Z}_{2},d\nu_{\mu}\right)  }\left(  \xi_{n}^{\mu
}\right)  $, $n=0,1,2,...$.

\bigskip

\section{Preliminary results}

\bigskip

In the calculations we will do in the Sections 4 and 5, the derivative of the
gamma function will arise naturally. Recall that the \textit{logarithmic
derivative} of $z\mapsto\Gamma\left(  z\right)  $, also called the
\textit{digamma function} and denoted by $\psi\left(  z\right)  $, is defined by%
\[
\psi\left(  z\right)  :=\frac{\Gamma^{\prime}\left(  z\right)  }{\Gamma\left(
z\right)  }%
\]
for all $z\neq0,-1,-2,....$ (See [Leb], p. 5.) We will be interested only in
the values and behavior of $\psi\left(  x\right)  $ with $x\in\mathbb{R}^{+}$.

From the basic property of the gamma function $\Gamma\left(  x+1\right)
=x\Gamma\left(  x\right)  $ one obtains the formula%
\[
\psi\left(  x+1\right)  =\frac{1}{x}+\psi\left(  x\right)  ,
\]
from which one gets by induction that%
\[
\psi\left(  x+n\right)  =\sum_{k=0}^{n-1}\frac{1}{x+k}+\psi\left(  x\right)
\]
for $n\in\mathbb{N}$. Using the identities $\psi\left(  1\right)  =-\gamma$
and $\psi\left(  \frac{1}{2}\right)  =-\gamma-2\log2$ (see [Leb], p. 6), the
previous formula implies (by taking $x=1$ and $x=\frac{1}{2}$) that%
\[
\psi\left(  n+1\right)  =-\gamma+\sum_{k=1}^{n}\frac{1}{k}%
\]
and%
\[
\psi\left(  n+\frac{1}{2}\right)  =-\gamma-2\log2+2\sum_{k=1}^{n}\frac
{1}{2k-1},
\]

When necessary we will use these formulas without further comment.

In this section we will state and prove two lemmas that we will be using in
Sections 4 and 5.

\bigskip

\textbf{Lemma 3.1 }
\textit{(a) The inequality }$0<\psi\left(  x+m\right)  -\log x<\left(
2m-1\right)  \left(  2x\right)  ^{-1}$\textit{\ holds for all }$x\in
\mathbb{R}^{+}$ \textit{and }$m\in\mathbb{N}$. \textit{In particular, we have
that for any }$m\in\mathbb{N}$%
\[
\lim_{x\rightarrow+\infty}\left(  \psi\left(  x+m\right)  -\log x\right)  =0.
\]
\textit{(b)} \textit{For }$y>0$\textit{\ fixed we have that }%
\[
\lim_{x\rightarrow+\infty}\left(  \psi\left(  x+y\right)  -\log x\right)  =0.
\]
\textit{(c) The inequality }$-x^{-1}<\psi\left(  x\right)  -\log x<-\left(
2x\right)  ^{-1}$\textit{\ holds for all} $x\in\mathbb{R}^{+}$. \textit{In
particular, we have that }%
\[
\lim_{x\rightarrow+\infty}\left(  \psi\left(  x\right)  -\log x\right)  =0.
\]

\bigskip

\textbf{Proof: }From the integral representation of $\psi\left(  z\right)  $,%
\[
\psi\left(  z\right)  =\int_{0}^{\infty}\left(  \frac{e^{-t}}{t}-\frac
{e^{-tz}}{1-e^{-t}}\right)  dt,
\]
and the integral representation of $\log\left(  z\right)  $,%
\[
\log\left(  z\right)  =\int_{0}^{\infty}\frac{e^{-t}-e^{-tz}}{t}dt,
\]
both valid for $\operatorname{Re}z>0$ (see [Leb], pp. 6,7), one obtains for
all $x>0$ and $m>0$ that%
\begin{equation}
\psi\left(  x+m\right)  -\log x=\int_{0}^{\infty}\left(  \frac{1}{t}%
-\frac{e^{-tm}}{1-e^{-t}}\right)  e^{-tx}dt.\tag{3.1}%
\end{equation}

For $m\in\mathbb{N}$, let us consider the function $h_{m}:\mathbb{R}%
\rightarrow\mathbb{R}$,%
\[
h_{m}\left(  t\right)  =\frac{1}{t}-\frac{e^{-tm}}{1-e^{-t}},
\]
where we define $h_{m}\left(  0\right)  =\lim_{t\rightarrow0}h_{m}\left(
t\right)  =\frac{2m-1}{2}>0$. So $h_{m}$ is continuous. For all $t>0$ we will
prove by induction that $0<h_{m}\left(  t\right)  <\frac{2m-1}{2}$ holds for
all $m\in\mathbb{N}$. Observe that $e^{t}>1+t$ for $t>0$ implies $h_{1}\left(
t\right)  >0$ for $t>0$. Also observe that $\beta\left(  t\right)  =\tanh
\frac{t}{2}-\frac{t}{2}$ is a decreasing function in $\mathbb{R}^{+}$, so that
$\tanh\frac{t}{2}<\frac{t}{2}$ for $t>0$, which implies that $h_{1}\left(
t\right)  <\frac{1}{2}$ for $t>0$. This proves the inequality $0<h_{m}\left(
t\right)  <\frac{2m-1}{2}$ for $m=1$. Suppose now that the inequality holds
for a given $m\in\mathbb{N}$. The hypothesis $h_{m}\left(  t\right)  >0$ gives us%
\[
h_{m+1}\left(  t\right)  =\frac{1}{t}-\frac{e^{-tm}}{1-e^{-t}}e^{-t}=\left(
\frac{1}{t}-\frac{e^{-tm}}{1-e^{-t}}\right)  e^{-t}+\frac{1-e^{-t}}{t}>0
\]
for $t>0$. Also, the case $m=1$ gives us that $\frac{1}{t}<\frac{1}{2}%
+\frac{e^{-t}}{1-e^{-t}}$, which together with the hypothesis $h_{m}\left(
t\right)  <\frac{2m-1}{2}$ gives us (for $t>0$) that
\begin{align*}
h_{m+1}\left(  t\right)   & =\left(  \frac{1}{t}-\frac{e^{-tm}}{1-e^{-t}%
}\right)  e^{-t}+\frac{1-e^{-t}}{t}\\
& <\frac{2m-1}{2}e^{-t}+\left(  1-e^{-t}\right)  \left(  \frac{1}{2}%
+\frac{e^{-t}}{1-e^{-t}}\right) \\
& =\frac{2m-1}{2}e^{-t}+\frac{1+e^{-t}}{2}\\
& =me^{-t}+\frac{1}{2}\\
& <\frac{2m+1}{2},
\end{align*}
as wanted. Then (3.1) and the inequality $0<h_{m}\left(  t\right)
<\frac{2m-1}{2}$ we just proved above gives us that%
\[
0<\psi\left(  x+m\right)  -\log x<\frac{2m-1}{2}\int_{0}^{\infty}%
e^{-tx}dt=\left(  2m-1\right)  \left(  2x\right)  ^{-1},
\]
which proves (a).

For $x\in\mathbb{R}^{+}$ we have that%
\[
\psi\left(  x\right)  -\log\left(  x\right)  =\psi\left(  x+1\right)
-\log\left(  x\right)  -x^{-1}.
\]

So, by using (a) with $m=1$ we have that%
\[
-x^{-1}<\psi\left(  x\right)  -\log\left(  x\right)  <\left(  2x\right)
^{-1}-x^{-1}=-\left(  2x\right)  ^{-1},
\]
which proves (c).

Now we prove (b). (We need to prove the result for $y\notin\mathbb{N}$.)
Observe that it is sufficient to demonstrate the result for $y\in\left(
0,1\right)  $, since given that for any fixed non-integer $Y>0$ we can write
$Y=\left\lfloor Y\right\rfloor +y$, where $\lfloor Y\rfloor$ is the floor
function of $Y$ and $y\in\left(  0,1\right)  $. Then, by defining
$X:=x+\left\lfloor Y\right\rfloor $ we have that%
\begin{align*}
\lim_{x\rightarrow+\infty}\left(  \psi\left(  x+Y\right)  -\log x\right)   &
=\lim_{X\rightarrow+\infty}\left(  \psi\left(  X+y\right)  -\log\left(
X-\left\lfloor Y\right\rfloor \right)  \right) \\
& =\lim_{X\rightarrow+\infty}\left(  \psi\left(  X+y\right)  -\log X-\log
\frac{X-\left\lfloor Y\right\rfloor }{X}\right) \\
& =\lim_{X\rightarrow+\infty}\left(  \psi\left(  X+y\right)  -\log X\right) \\
& =0.
\end{align*}

We consider the continuous function $h_{y}:\mathbb{R}\rightarrow\mathbb{R}$,
$h_{y}\left(  t\right)  =\frac{1}{t}-\frac{e^{-ty}}{1-e^{-t}}$, where
$h_{y}\left(  0\right)  =\lim_{t\rightarrow0}h_{y}\left(  t\right)
=\frac{2y-1}{2}$, and $0<y<1$ is fixed. According to (3.1), with $m=y\in
(0,1)$, it is sufficient to prove that $h_{y}$ is bounded in $\left[
0,\infty\right)  $, since if $\left\vert h_{y}\left(  t\right)  \right\vert
\leq C$ for all $t\geq0$, then%
\[
\left\vert \psi\left(  x+y\right)  -\log x\right\vert =\left\vert \int
_{0}^{\infty}h_{y}\left(  t\right)  e^{-tx}dt\right\vert \leq C\int
_{0}^{\infty}e^{-tx}dt=\frac{C}{x},
\]
and thus $\psi\left(  x+y\right)  -\log x\rightarrow0$ as $x\rightarrow
+\infty$. But observe that $\lim_{t\rightarrow+\infty}h_{y}\left(  t\right)
=0$ and that $h_{y}$ is continuous, which shows that $h_{y}$ is bounded on
$\left[  0,\infty\right)  $.

\hfill\textbf{Q.E.D.}

\bigskip

\textbf{Lemma 3.2 }\textit{Let }$\mu>-\frac{1}{2}$\textit{\ be fixed. Then }
\[
\lim_{n\rightarrow\infty}\frac{\left(  \gamma_{\mu}\left(  n\right)  \right)
^{\frac{1}{n}}}{n}=e^{-1}.
\]

\textit{(Note that this limit does not depend on }$\mu$\textit{.)}

\bigskip

\textbf{Proof: \ }It is sufficient to prove that%
\[
\lim_{n\rightarrow\infty}\frac{\left(  \gamma_{\mu}\left(  2n\right)  \right)
^{\frac{1}{2n}}}{2n}=\lim_{n\rightarrow\infty}\frac{\left(  \gamma_{\mu
}\left(  2n+1\right)  \right)  ^{\frac{1}{2n+1}}}{2n+1}=e^{-1}.
\]

Let us consider the even case. We can write by using formula (2.1) that%
\[
\frac{\left(  \gamma_{\mu}(2n)\right)  ^{\frac{1}{2n}}}{2n}=\frac{\left(
\left(  2n\right)  !\right)  ^{\frac{1}{2n}}}{2n}\left(  \frac{\Gamma\left(
\frac{1}{2}\right)  }{\Gamma\left(  \mu+\frac{1}{2}\right)  }\right)
^{\frac{1}{2n}}\left(  \frac{\Gamma\left(  \mu+n+\frac{1}{2}\right)  }%
{\Gamma\left(  n+\frac{1}{2}\right)  }\right)  ^{\frac{1}{2n}}.
\]

We have that $\lim_{n\rightarrow\infty}\frac{\left(  \left(  2n\right)
!\right)  ^{\frac{1}{2n}}}{2n}=e^{-1}$ and $\lim_{n\rightarrow\infty}\left(
\frac{\Gamma\left(  \frac{1}{2}\right)  }{\Gamma\left(  \mu+\frac{1}%
{2}\right)  }\right)  ^{\frac{1}{2n}}=1$. So it remains to prove that the
limit of the third factor in the left hand side is $1$. By using Stirling's
formula we have that%
\begin{align*}
\lim_{n\rightarrow\infty}\left(  \frac{\Gamma\left(  \mu+n+\frac{1}{2}\right)
}{\Gamma\left(  n+\frac{1}{2}\right)  }\right)  ^{\frac{1}{2n}}  &
=\lim_{n\rightarrow\infty}\left(  \frac{\sqrt{2\pi}\left(  \mu+n+\frac{1}%
{2}\right)  ^{\mu+n}e^{-\left(  \mu+n+\frac{1}{2}\right)  }}{\sqrt{2\pi
}\left(  n+\frac{1}{2}\right)  ^{n}e^{-\left(  n+\frac{1}{2}\right)  }%
}\right)  ^{\frac{1}{2n}}\\
& =\lim_{n\rightarrow\infty}\left(  \left(  \mu+n+\frac{1}{2}\right)
^{\frac{\mu}{2n}}e^{-\frac{\mu}{2n}}\left(  \frac{\mu+n+\frac{1}{2}}%
{n+\frac{1}{2}}\right)  ^{\frac{1}{2}}\right) \\
& =1.
\end{align*}

For the odd case, by using (2.2) we have that%
\[
\frac{\left(  \gamma_{\mu}(2n+1)\right)  ^{\frac{1}{2n+1}}}{2n+1}%
=\frac{\left(  \left(  2n+1\right)  !\right)  ^{\frac{1}{2n+1}}}{2n+1}\left(
\frac{\Gamma\left(  \frac{1}{2}\right)  }{\Gamma\left(  \mu+\frac{1}%
{2}\right)  }\right)  ^{\frac{1}{2n+1}}\left(  \frac{\Gamma\left(  \mu
+n+\frac{3}{2}\right)  }{\Gamma\left(  n+\frac{3}{2}\right)  }\right)
^{\frac{1}{2n+1}}.
\]

We have that $\lim_{n\rightarrow\infty}\frac{\left(  \left(  2n+1\right)
!\right)  ^{\frac{1}{2n+1}}}{2n+1}=e^{-1}$ and $\lim_{n\rightarrow\infty
}\left(  \frac{\Gamma\left(  \frac{1}{2}\right)  }{\Gamma\left(  \mu+\frac
{1}{2}\right)  }\right)  ^{\frac{1}{2n+1}}=1$. So the proof ends by showing
that the limit of the third factor in the left hand side is $1$. By using
Stirling's formula we have that%
\begin{align*}
& \lim_{n\rightarrow\infty}\left(  \frac{\Gamma\left(  \mu+n+\frac{3}%
{2}\right)  }{\Gamma\left(  n+\frac{3}{2}\right)  }\right)  ^{\frac{1}{2n+1}%
}\\
& =\lim_{n\rightarrow\infty}\left(  \frac{\sqrt{2\pi}\left(  \mu+n+\frac{3}%
{2}\right)  ^{\mu+n+1}e^{-\left(  \mu+n+\frac{3}{2}\right)  }}{\sqrt{2\pi
}\left(  n+\frac{3}{2}\right)  ^{n+1}e^{-\left(  n+\frac{3}{2}\right)  }%
}\right)  ^{\frac{1}{2n+1}}\\
& =\lim_{n\rightarrow\infty}\left(  \left(  \mu+n+\frac{3}{2}\right)
^{\frac{\mu}{2n+1}}e^{-\frac{\mu}{2n+1}}\left(  \frac{\mu+n+\frac{3}{2}%
}{n+\frac{3}{2}}\right)  ^{\frac{n+1}{2n+1}}\right) \\
& =1.
\end{align*}

\hfill\textbf{Q.E.D.}

\bigskip

Observe that formula (1.1), which gives us the entropy of the elements of the
canonical basis $\left\{  \xi_{n}\right\}  $ of $\mathcal{B}^{2}$, can be
written as%
\begin{equation}
S_{L^{2}\left(  \mathbb{C},d\nu_{\text{Gauss}}\right)  }\left(  \xi
_{n}\right)  =n\psi\left(  n+1\right)  -\log n!.\tag{3.2}%
\end{equation}

In the case $n=0$ we have $\xi_{0}=1$ and then from (2.15) we have that
$S_{L^{2}\left(  \mathbb{C},d\nu_{\text{Gauss}}\right)  }\left(  1\right)
=0$. (Note that this case is also included in (3.2).)

We can use Lemmas 3.1 and 3.2 to prove some properties of the sequence of
entropies $\left\{  S_{n}\right\}  _{n=0}^{\infty}$, where $S_{n}%
:=S_{L^{2}\left(  \mathbb{C},d\nu_{\text{Gauss}}\right)  }\left(  \xi
_{n}\right)  $. First, we note that%
\begin{align*}
S_{n+1}  & =\left(  n+1\right)  \psi\left(  n+2\right)  -\log\left(
n+1\right)  !\\
& =\left(  n+1\right)  \left(  \frac{1}{n+1}+\psi\left(  n+1\right)  \right)
-\log n!-\log\left(  n+1\right) \\
& =S_{n}+1+\psi\left(  n+1\right)  -\log\left(  n+1\right) \\
& >S_{n}+1-\frac{1}{n+1},
\end{align*}
where we used Lemma 3.1 (c). Thus, for $n=0$ we have that $S_{1}>0$, and for
$n\in\mathbb{N}$ we have $S_{n+1}>S_{n}$. That is, the sequence $\left\{
S_{n}\right\}  _{n=0}^{\infty}$ is increasing. Moreover, $\left\{
S_{n}\right\}  _{n=0}^{\infty}$ is a sequence of non-negative terms. (This
conclusion also comes from the fact that $\left(  \mathbb{C},d\nu
_{\text{Gauss}}\right)  $ is a probability measure space.)

Next, by using the equality $S_{n+1}-S_{n}=1+\psi\left(  n+1\right)
-\log\left(  n+1\right)  $ of the previous argument and Lemma (3.1) (c) we
have that%
\[
\lim_{n\rightarrow\infty}\left(  S_{n+1}-S_{n}\right)  =1,
\]
which proves that the sequence $\left\{  S_{n}\right\}  _{n=1}^{\infty}$ is
unbounded and, moreover, implies that%
\[
\lim_{n\rightarrow\infty}\frac{S_{n}}{n}=1.
\]
(Proof: $\lim_{n\rightarrow\infty}\left(  S_{n+1}-S_{n}\right)  =1\Rightarrow
\lim_{n\rightarrow\infty}\frac{1}{n}\sum\limits_{k=0}^{n-1}\left(
S_{k+1}-S_{k}\right)  =1 \linebreak \Rightarrow\lim_{n\rightarrow\infty}\frac{S_{n}-S_{0}%
}{n}=1\Rightarrow\lim_{n\rightarrow\infty}\frac{S_{n}}{n}=1$.) This limit can
also be proved directly by noting that%
\begin{align*}
\frac{S_{n}}{n}  & =\psi\left(  n+1\right)  -\frac{1}{n}\log n!\\
& =\psi\left(  n+1\right)  -\log n-\log\frac{\left(  n!\right)  ^{\frac{1}{n}%
}}{n},
\end{align*}
and thus, by using that $\psi\left(  n+1\right)  -\log n\rightarrow0$ as
$n\rightarrow\infty$ (Lemma 3.1 (a)) and that $\log\frac{\left(  n!\right)
^{\frac{1}{n}}}{n}\rightarrow e^{-1}$ as $n\rightarrow\infty$, we obtain the
desired result $\lim_{n\rightarrow\infty}\frac{S_{n}}{n}=1\,.$

In conclusion, we have proved the following.

\bigskip

\textbf{Proposition 3.1 \ }\textit{The sequence }$\left\{  S_{n}\right\}
_{n=0}^{\infty}$\textit{, where } $S_{n}=S_{L^{2}\left(  \mathbb{C}%
,d\nu_{\text{Gauss}}\right)  }\left(  \xi_{n}\right)  $\textit{\ is the
entropy of the $n$-th canonical basis element in }${L^{2}\left(
\mathbb{C},d\nu_{\text{Gauss}}\right)  }$ \textit{is an\linebreak unbounded
increasing sequence of non-negative terms, with the property }\linebreak%
$\lim_{n\rightarrow\infty}\left(  S_{n+1}-S_{n}\right)  =1$\textit{\ (which
implies that }$\lim_{n\rightarrow\infty}\frac{S_{n}}{n}=1$\textit{).}

\bigskip

\section{Entropies in $\mathcal{B}_{\mu}^{2}$}

\bigskip

As noted in Section 2, for calculating the entropies $S_{n}^{\mu}%
=S_{L^{2}\left(  \mathbb{C}\times\mathbb{Z}_{2},d\nu_{\mu}\right)  }\left(
\xi_{n}^{\mu}\right)  $ of the elements of the canonical basis $\left\{
\xi_{n}^{\mu}\right\}  _{n=0}^{\infty}$ of the $\mu$-deformed Segal-Bargmann
space $\mathcal{B}_{\mu}^{2}$, we need to consider the cases when $n$ is even
(in which case we have that $S_{n}^{\mu}=S_{L^{2}\left(  \mathbb{C}%
,d\nu_{e,\mu}\right)  }\left(  \xi_{n}\right)  $) and when $n$ is odd (in
which case we have that $S_{n}^{\mu}=S_{L^{2}\left(  \mathbb{C},d\nu_{o,\mu
}\right)  }\left(  \xi_{n}\right)  $). We begin by considering the even case.
For $n=0$ we have $\xi_{0}^{\mu}\left(  z\right)  =1$ and then $S_{0}^{\mu}%
=0$. So we are interested in calculating $S_{2n}^{\mu}$ for $n\geq1$. Formula
(2.15) tells us that%
\begin{align*}
S_{2n}^{\mu}  & =\int_{\mathbb{C}}\left\vert \xi_{2n}\left(  z\right)
\right\vert ^{2}\!\log\left\vert \xi_{2n}\left(  z\right)  \right\vert
^{2}\!d\nu_{e,\mu}\left(  z\right)  \!-\!\left\Vert \xi_{2n}\right\Vert
_{L^{2}\left(  \mathbb{C},d\nu_{e,\mu}\right)  }^{2}\!\log\left\Vert \xi
_{2n}\right\Vert _{L^{2}\left(  \mathbb{C},d\nu_{e,\mu}\right)  }^{2}\\
& =\frac{2^{\frac{1}{2}-\mu}}{\pi\Gamma\left(  \mu+\frac{1}{2}\right)  }%
\!\int_{\mathbb{C}}\left\vert \frac{z^{2n}}{\left(  \gamma_{\mu}(2n)\right)
^{\frac{1}{2}}}\right\vert ^{2}\!\!\log\left\vert \frac{z^{2n}}{\left(
\gamma_{\mu}(2n)\right)  ^{\frac{1}{2}}}\right\vert ^{2}\!\!K_{\mu-\frac{1}%
{2}}( \left\vert z\right\vert ^{2}) \left\vert z\right\vert ^{2\mu+1}\!dxdy.
\end{align*}

Since the $\log$ term in the integral of the right hand side is $\log
\left\vert z^{2n}\right\vert ^{2}-\log\gamma_{\mu}(2n)$, we can write
$S_{2n}^{\mu}$ as a difference of two integrals, $I_{1}-I_{2}$ say, in which
$I_{2}=\log\gamma_{\mu}(2n)\left\Vert \xi_{2n}\right\Vert _{L^{2}\left(
\mathbb{C},d\nu_{e,\mu}\right)  }^{2}=\log\gamma_{\mu}(2n)$. In $I_{1}$ we
change $\left(  x,y\right)  $ to polar coordinates $\left(  r,\theta\right)
$, and then let $s=r^{2}$ to obtain%
\begin{align*}
S_{2n}^{\mu}  & =\frac{2^{\frac{1}{2}-\mu}}{\pi\Gamma\left(  \mu+\frac{1}%
{2}\right)  }\int_{\mathbb{C}}\left\vert \frac{z^{2n}}{\left(  \gamma_{\mu
}(2n)\right)  ^{\frac{1}{2}}}\right\vert ^{2}\log\left\vert z^{2n}\right\vert
^{2}K_{\mu-\frac{1}{2}}\left(  \left\vert z\right\vert ^{2}\right)  \left\vert
z\right\vert ^{2\mu+1}dxdy\\
& -\log\gamma_{\mu}(2n)\\
& =\frac{2^{\frac{1}{2}-\mu}2}{\gamma_{\mu}(2n)\Gamma\left(  \mu+\frac{1}%
{2}\right)  }\int_{0}^{\infty}r^{4n}\left(  \log r^{4n}\right)  K_{\mu
-\frac{1}{2}}\left(  r^{2}\right)  r^{2\mu+2}dr-\log\gamma_{\mu}(2n)\\
& =\frac{2^{\frac{1}{2}-\mu}}{\gamma_{\mu}(2n)\Gamma\left(  \mu+\frac{1}%
{2}\right)  }\int_{0}^{\infty}s^{2n}\left(  \log s^{2n}\right)  K_{\mu
-\frac{1}{2}}\left(  s\right)  s^{\mu+\frac{1}{2}}ds-\log\gamma_{\mu}(2n).
\end{align*}

For calculating the integral $\int_{0}^{\infty}K_{\mu-\frac{1}{2}}\left(
s\right)  s^{\mu+2n+\frac{1}{2}}ds$, we define the function $\varphi$ in a
neighborhood of $\alpha=1$ as%
\[
\varphi\left(  \alpha\right)  =\int_{0}^{\infty}s^{2n\alpha}K_{\mu-\frac{1}%
{2}}\left(  s\right)  s^{\mu+\frac{1}{2}}ds.
\]

Observe that for $\mu>-\frac{1}{2}$, $n\in\mathbb{N}$ and $\alpha$ in a
neighborhood of $1$, one has that $2n\alpha+\mu+\frac{3}{2}>\left\vert
\mu-\frac{1}{2}\right\vert $, so we can use formula (2.7) to write%
\[
\varphi\left(  \alpha\right)  =2^{2n\alpha+\mu-\frac{1}{2}}\Gamma\left(
n\alpha+1\right)  \Gamma\left(  \mu+n\alpha+\frac{1}{2}\right)  .
\]

The derivative $\varphi^{\prime}$ is on the one hand%
\[
\varphi^{\prime}\left(  \alpha\right)  =\int_{0}^{\infty}s^{2n\alpha}\left(
\log s^{2n}\right)  K_{\mu-\frac{1}{2}}\left(  s\right)  s^{\mu+\frac{1}{2}%
}ds,
\]
and on the other hand%
\begin{align*}
\varphi^{\prime}\left(  \alpha\right)   & =2^{2n\alpha+\mu-\frac{1}{2}}%
\Gamma\left(  n\alpha+1\right)  n\Gamma^{\prime}\left(  \mu+n\alpha+\frac
{1}{2}\right) \\
& +2^{2n\alpha+\mu-\frac{1}{2}}n\Gamma^{\prime}\left(  n\alpha+1\right)
\Gamma\left(  \mu+n\alpha+\frac{1}{2}\right) \\
& +2^{2n\alpha+\mu-\frac{1}{2}}2n(\log2)\Gamma\left(  n\alpha+1\right)
\Gamma\left(  \mu+n\alpha+\frac{1}{2}\right) \\
& =2^{2n\alpha+\mu-\frac{1}{2}}\Gamma\left(  n\alpha+1\right)  \Gamma\left(
\mu+n\alpha+\frac{1}{2}\right)  \left(
\begin{array}
[c]{c}%
n\psi\left(  \mu+n\alpha+\frac{1}{2}\right) \\
+n\psi\left(  n\alpha+1\right) \\
+2n\log2
\end{array}
\right)  .
\end{align*}

Then%
\begin{align*}
\varphi^{\prime}\left(  1\right)   & =\int_{0}^{\infty}s^{2n}\left(  \log
s^{2n}\right)  K_{\mu-\frac{1}{2}}\left(  s\right)  s^{\mu+\frac{1}{2}}ds\\
& =2^{2n+\mu-\frac{1}{2}}\Gamma\left(  n+1\right)  \Gamma\left(  \mu
+n+\frac{1}{2}\right)  \left(
\begin{array}
[c]{c}%
n\psi\left(  \mu+n+\frac{1}{2}\right) \\
+n\psi\left(  n+1\right) \\
+2n\log2
\end{array}
\right)  .
\end{align*}

Thus we have that%
\begin{align*}
S_{2n}^{\mu} \!\! & =\!\!\frac{\Gamma\left(  n+1\right)  \Gamma\left(
\mu+n+\frac{1}{2}\right)  2^{2n}}{\gamma_{\mu}(2n)\Gamma\left(  \mu+\frac
{1}{2}\right)  }\left( \! n\psi\left(  \!\mu\!+\!n\!+\!\frac{1}{2}\!\right)
+n\psi\left(  n+1\right)  +\log2^{2n}\!\right) \\
& -\log\gamma_{\mu}(2n).
\end{align*}

By using formula (2.1) for $\gamma_{\mu}\left(  2n\right)  $ we have that the
entropy of the even elements $\xi_{2n}$ is%
\begin{equation}
S_{2n}^{\mu}=n\left(  \psi\left(  \mu+n+\frac{1}{2}\right)  +\psi\left(
n+1\right)  \right)  -\log\frac{\gamma_{\mu}(2n)}{2^{2n}}.\tag{4.1}%
\end{equation}

Note that this formula makes sense for $n=0$, obtaining the known result
$S_{0}^{\mu}=0$.

In the case $\mu=0$, formula (4.1) becomes%
\begin{align*}
S_{2n}^{0}  & =n\left(  \psi\left(  n+\frac{1}{2}\right)  +\psi\left(
n+1\right)  \right)  -\log\frac{(2n)!}{2^{2n}}\\
& =n\left(  -\gamma-2\log2+2\sum_{k=1}^{n}\frac{1}{2k-1}-\gamma+\sum_{k=1}%
^{n}\frac{1}{k}\right)  -\log\left(  2n\right)  !+2n\log2\\
& =2n\left(  -\gamma+\sum_{k=1}^{n}\frac{1}{2k-1}+\frac{1}{2}\sum_{k=1}%
^{n}\frac{1}{k}\right)  -\log\left(  2n\right)  !\\
& =2n\left(  -\gamma+\sum_{k=1}^{2n}\frac{1}{k}\right)  -\log\left(
2n\right)  !,
\end{align*}
which is (1.1) for even positive integers, as expected.

Since $\left(  \mathbb{C},d\nu_{e,\mu}\right)  $ is a probability measure
space, we have that $S_{2n}^{\mu}\geq0$ for all $n=0,1,2,...$. But we can
arrive at this conclusion directly from the formula obtained for $S_{2n}^{\mu
}$ as follows. Observe that for $n\in\mathbb{N}$ we can write formula (2.1) as%
\begin{equation}
\frac{\gamma_{\mu}(2n)}{2^{2n}}=n!\prod_{k=1}^{n}\left(  \mu+k-\frac{1}%
{2}\right)  .\tag{4.2}%
\end{equation}

Then%
\begin{align*}
S_{2n}^{\mu}  & =n\psi\left(  \mu+n+\frac{1}{2}\right)  +n\psi\left(
n+1\right)  -\log\left(  n!\prod_{k=1}^{n}\left(  \mu+k-\frac{1}{2}\right)
\right) \\
& =\sum_{k=1}^{n}\left(  \psi\left(  \mu+n+\frac{1}{2}\right)  -\log\left(
\mu+k-\frac{1}{2}\right)  +\psi\left(  n+1\right)  -\log\left(  k\right)
\right)  .
\end{align*}

Lemma 3.1(a) gives us that $\psi\left(  \mu+n+\frac{1}{2}\right)  -\log\left(
\mu+k-\frac{1}{2}\right)  >0$ and that $\psi\left(  n+1\right)  -\log\left(
k\right)  >0$ for all $k=1,...,n$. So we conclude that $S_{2n}^{\mu}>0 $, as
wanted. Moreover, observe that for fixed $n\in\mathbb{N}$, we have that (again
by Lemma 3.1(a)) $\psi\left(  \mu+n+\frac{1}{2}\right)  -\log\left(
\mu+k-\frac{1}{2}\right)  \rightarrow0$ as $\mu\rightarrow+\infty$, and so%
\[
\lim_{\mu\rightarrow+\infty}S_{2n}^{\mu}=\sum_{k=1}^{n}\left(  \psi\left(
n+1\right)  -\log\left(  k\right)  \right)  =n\psi\left(  n+1\right)  -\log
n!.
\]

That is, for $n\in\mathbb{N}$ fixed we have that%
\[
\lim_{\mu\rightarrow+\infty}S_{2n}^{\mu}=S_{n}.
\]

Let us consider the particular case when $\mu=\frac{1}{2}+m$, $m=0,1,2,...$.
Formula (4.2) becomes in this case%
\[
\frac{\gamma_{\frac{1}{2}+m}(2n)}{2^{2n}}=n!\prod_{k=1}^{n}\left(  k+m\right)
=\frac{n!\left(  m+n\right)  !}{m!},
\]
and then formula (4.1) gives us
\begin{align*}
S_{2n}^{\frac{1}{2}+m}  & =n\left(  \psi\left(  n+m+1\right)  +\psi\left(
n+1\right)  \right)  -\log\frac{n!\left(  m+n\right)  !}{m!}\\
& =\left(  n+m\right)  \psi\left(  n+m+1\right)  -\log\left(  m+n\right)
!+n\psi\left(  n+1\right) \\
& -\log n!-m\psi\left(  n+m+1\right)  +\log m!\\
& =S_{n+m}+S_{n}-m\left(  \sum_{k=0}^{n-1}\frac{1}{m+k+1}+\psi\left(
m+1\right)  \right)  +\log m!\\
& =S_{n+m}+S_{n}-S_{m}-\sum_{k=1}^{n}\frac{m}{m+k}.
\end{align*}

That is, for $n,m=0,1,2,...$, we have the formula%
\[
S_{n+m}+S_{n}-S_{m}=S_{2n}^{\frac{1}{2}+m}+\sum_{k=1}^{n}\frac{m}{m+k},
\]
which shows that the values of the entropies $S_{n+m},S_{n}$ and $S_{m}$ (of
the undeformed case) are related by means of the entropy $S_{2n}^{\frac{1}%
{2}+m}$ corresponding to the $\left(  m+\frac{1}{2}\right)  $-deformed case.

We claim that $\left\{  S_{2n}^{\mu}\right\}  _{n=0}^{\infty}$ is an
increasing sequence for fixed $\mu>-\frac{1}{2}$. In fact, we have that%
\begin{align*}
S_{2n+2}^{\mu}  & =\left(  n+1\right)  \psi\left(  \mu+n+\frac{3}{2}\right)
+\left(  n+1\right)  \psi\left(  n+2\right)  -\log\frac{\gamma_{\mu}%
(2n+2)}{2^{2n+2}}\\
& =n\left(  \frac{1}{\mu+n+\frac{1}{2}}+\psi\left(  \mu+n+\frac{1}{2}\right)
\right)  +\psi\left(  \mu+n+\frac{3}{2}\right) \\
& +n\left(  \frac{1}{n+1}+\psi\left(  n+1\right)  \right)  +\psi\left(
n+2\right) \\
& -\log\frac{\left(  2n+2\right)  \left(  2n+1+2\mu\right)  \gamma_{\mu}%
(2n)}{2^{2}2^{2n}}\\
& =S_{2n}^{\mu}+\psi\left(  \mu+n+\frac{3}{2}\right)  -\log\left(  \mu
+n+\frac{1}{2}\right) \\
& +\psi\left(  n+2\right)  -\log\left(  n+1\right)  +\frac{n}{\mu+n+\frac
{1}{2}}+\frac{n}{n+1}.
\end{align*}

Lemma 3.1(a) gives us $\psi\left(  \mu+n+\frac{3}{2}\right)  -\log\left(
\mu+n+\frac{1}{2}\right)  >0$ and $\psi\left(  n+2\right)  -\log\left(
n+1\right)  >0$. Thus we have that $S_{2n+2}^{\mu}-S_{2n}^{\mu}>0$, as wanted.
Lemma 3.1(a) also tells us that $\psi\left(  \mu+n+\frac{3}{2}\right)
-\log\left(  \mu+n+\frac{1}{2}\right)  \rightarrow0$ and $\psi\left(
n+2\right)  -\log\left(  n+1\right)  \rightarrow0$ as $n\rightarrow\infty$.
Thus, for fixed $\mu>-\frac{1}{2}$, we have by the expression above that
$\lim_{n\rightarrow\infty}\left(  S_{2n+2}^{\mu}-S_{2n}^{\mu}\right)  =2$. In
particular we see that the sequence $\left\{  S_{2n}^{\mu}\right\}
_{n=0}^{\infty}$ is unbounded. This limit implies that $\lim_{n\rightarrow
\infty}\frac{S_{2n}}{2n}=1$, but we can give a direct proof of this last
result by noting that%
\begin{align*}
\frac{S_{2n}^{\mu}}{2n}  & =\frac{1}{2}\left(  \psi\left(  \mu+n+\frac{1}%
{2}\right)  +\psi\left(  n+1\right)  \right)  -\frac{1}{2n}\log\frac
{\gamma_{\mu}(2n)}{2^{2n}}\\
& =\frac{1}{2}\left(  \psi\left(  \mu+n+\frac{1}{2}\right)  +\psi\left(
n+1\right)  \right)  -\log\frac{\left(  \gamma_{\mu}(2n)\right)  ^{\frac
{1}{2n}}}{2n}-\log n\\
& =\frac{1}{2}\left(  \psi\left(  \mu+n+\frac{1}{2}\right)  -\log
n+\psi\left(  n+1\right)  -\log n\right)  -\log\frac{\left(  \gamma_{\mu
}(2n)\right)  ^{\frac{1}{2n}}}{2n}.
\end{align*}

Lemma 3.1(b) tells us that $\psi\left(  \mu+n+\frac{1}{2}\right)  -\log
n\rightarrow0$ and $\psi\left(  n+1\right)  -\log n\rightarrow0$ as
$n\rightarrow\infty$. Lemma 3.2 tells us that $\log\frac{\left(  \gamma_{\mu
}(2n)\right)  ^{\frac{1}{2n}}}{2n}\rightarrow-1$ as $n\rightarrow\infty$. Then
we have that $\lim_{n\rightarrow\infty}\frac{S_{2n}}{2n}\rightarrow1$, as wanted.

In conclusion, we have proved the following theorem.

\bigskip

\textbf{Theorem 4.1 }\textit{The entropy of }$\xi_{2n}^{\mu}$\textit{\ is
given by}%
\[
S_{2n}^{\mu}=n\left(  \psi\left(  \mu+n+\frac{1}{2}\right)  +\psi\left(
n+1\right)  \right)  -\log\frac{\gamma_{\mu}(2n)}{2^{2n}},
\]
\textit{where }$\mu>-\frac{1}{2}$\textit{\ and }$n=0,1,...$\textit{. For fixed
}$\mu>-\frac{1}{2}$\textit{, the sequence }$\left\{  S_{2n}^{\mu}\right\}
_{n=1}^{\infty}$\textit{\ is an unbounded increasing sequence of positive
terms such that}%
\[
\lim_{n\rightarrow\infty}\left(  S_{2n+2}^{\mu}-S_{2n}^{\mu}\right)  =2,
\]
\textit{which implies that }
\[
\lim_{n\rightarrow\infty}\frac{S_{2n}^{\mu}}{2n}=1.
\]

\textit{For fixed }$n\in\mathbb{N}$\textit{, we have that}%
\[
\lim_{\mu\rightarrow+\infty}S_{2n}^{\mu}=S_{n},
\]
\textit{where }$S_{n}=S_{n}^{0}$\textit{.}

\textit{For }$n,m=0,1,2,...$\textit{, we have that}%
\[
S_{n+m}+S_{n}-S_{m}=S_{2n}^{\frac{1}{2}+m}+\sum_{k=0}^{n-1}\frac{m}{m+k+1}.
\]

\bigskip

We now calculate the entropies of the odd functions $\xi_{2n+1}$,
$n=0,1,2,...$. The steps we will follow in the calculations are analogues of
the even case. Since $S_{2n+1}^{\mu}=S_{L^{2}\left(  \mathbb{C},d\nu_{o,\mu
}\right)  }\left(  \xi_{2n+1}\right)  $ we have that%
\begin{align*}
S_{2n+1}^{\mu}  & =\frac{2^{\frac{1}{2}-\mu}}{\pi\Gamma\left(  \mu+\frac{1}%
{2}\right)  }\\
& \cdot\int_{\mathbb{C}}\left\vert \frac{z^{2n+1}}{\left(  \gamma_{\mu
}(2n\!+\!1)\right)  ^{\frac{1}{2}}}\right\vert ^{2}\log\left\vert
\frac{z^{2n+1}}{\left(  \gamma_{\mu}(2n\!+\!1)\right)  ^{\frac{1}{2}}%
}\right\vert ^{2}\!K_{\mu+\frac{1}{2}}\left(  \left\vert z\right\vert
^{2}\right)  \!\left\vert z\right\vert ^{2\mu+1}dxdy\\
& =\frac{2^{\frac{1}{2}-\mu}2}{\gamma_{\mu}(2n+1)\Gamma\left(  \mu+\frac{1}%
{2}\right)  }\int_{0}^{\infty}r^{4n+2}\left(  \log r^{4n+2}\right)
K_{\mu+\frac{1}{2}}\left(  r^{2}\right)  r^{2\mu+2}dr\\
& -\log\gamma_{\mu}(2n+1)\\
& =\frac{2^{\frac{1}{2}-\mu}}{\gamma_{\mu}(2n+1)\Gamma\left(  \mu+\frac{1}%
{2}\right)  }\int_{0}^{\infty}s^{2n+1}\left(  \log s^{2n+1}\right)
K_{\mu+\frac{1}{2}}\left(  s\right)  s^{\mu+\frac{1}{2}}ds\\
& -\log\gamma_{\mu}(2n+1).
\end{align*}

We define%
\[
\phi\left(  \alpha\right)  =\int_{0}^{\infty}s^{\left(  2n+1\right)  \alpha
}K_{\mu+\frac{1}{2}}\left(  s\right)  s^{\mu+\frac{1}{2}}ds.
\]

Since for $\mu>-\frac{1}{2}$, $n\in\mathbb{N\cup}\{0\}$ and $\alpha$ in a
neighborhood of $1$, one has that $\left(  2n+1\right)  \alpha+\mu+\frac{3}%
{2}>\left\vert \mu+\frac{1}{2}\right\vert $, we can use formula (2.7) to write%
\[
\phi\left(  \alpha\right)  =2^{\left(  2n+1\right)  \alpha+\mu-\frac{1}{2}%
}\Gamma\left(  \left(  n+\frac{1}{2}\right)  \alpha+\frac{1}{2}\right)
\Gamma\left(  \left(  n+\frac{1}{2}\right)  \alpha+\mu+1\right)  .
\]

By calculating the derivative $\phi^{\prime}(1)$ in two different ways as we
did in the even case, we get%
\begin{align*}
\phi^{\prime}\left(  1\right)   & =\int_{0}^{\infty}s^{2n+1}\log
s^{2n+1}K_{\mu+\frac{1}{2}}\left(  s\right)  s^{\mu+\frac{1}{2}}ds\\
& =2^{2n+\mu+\frac{1}{2}}\Gamma\left(  n+1\right)  \Gamma\left(  \mu
+n+\frac{3}{2}\right)  \left(
\begin{array}
[c]{c}%
\frac{2n+1}{2}\psi\left(  \mu+n+\frac{3}{2}\right) \\
+\frac{2n+1}{2}\psi\left(  n+1\right) \\
+(2n+1)\log2
\end{array}
\right)  .
\end{align*}

Thus, by using formula (2.2) for $\gamma_{\mu}(2n+1)$ we find that the entropy
of $\xi_{2n+1}$ is%
\begin{equation}
S_{2n+1}^{\mu}=\left(  n+\frac{1}{2}\right)  \left(  \psi\left(  \mu
+n+\frac{3}{2}\right)  +\psi\left(  n+1\right)  \right)  -\log\frac
{\gamma_{\mu}(2n+1)}{2^{2n+1}}.\tag{4.3}%
\end{equation}

In the case $\mu=0$ this formula becomes%
\begin{align*}
S_{2n+1}^{0}  & =\left(  n+\frac{1}{2}\right)  \left(  \psi\left(  n+\frac
{3}{2}\right)  +\psi\left(  n+1\right)  \right)  -\log\frac{(2n+1)!}{2^{2n+1}%
}\\
& =\left(  n+\frac{1}{2}\right)  \left(  \frac{1}{n+\frac{1}{2}}+\psi\left(
n+\frac{1}{2}\right)  +\psi\left(  n+1\right)  \right) \\
& -\log\left(  2n+1\right)  !+\left(  2n+1\right)  \log2\\
& =\left(  2n+1\right)  \left(  -\gamma+\frac{1}{2n+1}+\sum_{k=1}^{n}\frac
{1}{2k-1}+\frac{1}{2}\sum_{k=1}^{n}\frac{1}{k}\right)  -\log\left(
2n+1\right)  !\\
& =\left(  2n+1\right)  \left(  -\gamma+\sum_{k=1}^{2n+1}\frac{1}{k}\right)
-\log\left(  2n+1\right)  !,
\end{align*}
which is (1.1) for odd positive integers.

Observe that for $n\in\mathbb{N}$ we can write formula (2.2) as%
\[
\frac{\gamma_{\mu}(2n+1)}{2^{2n+1}}=n!\prod_{k=1}^{n+1}\left(  \mu+k-\frac
{1}{2}\right)  .
\]

Thus (4.3) can be written as%
\begin{align}
S_{2n+1}^{\mu}  & =\sum_{k=1}^{n}\left(  \psi\left(  \mu+n+\frac{3}{2}\right)
-\log\left(  \mu+k-\frac{1}{2}\right)  \right) \tag{4.4}\\
& +\frac{1}{2}\left(  \psi\left(  \mu+n+\frac{3}{2}\right)  -\log\left(
\mu+n+\frac{1}{2}\right)  \right) \nonumber\\
& +\left(  n+\frac{1}{2}\right)  \psi\left(  n+1\right)  -\log n!\nonumber\\
& -\frac{1}{2}\log\left(  \mu+n+\frac{1}{2}\right)  .\nonumber
\end{align}

For fixed $n=0,1,2,...$, we have by Lemma 3.1(a) that $\psi\left(  \mu
+n+\frac{3}{2}\right)  -\log\left(  \mu+k-\frac{1}{2}\right)  \rightarrow0$
and $\psi\left(  \mu+n+\frac{3}{2}\right)  -\log\left(  \mu+n+\frac{1}%
{2}\right)  \rightarrow0$ as $\mu\rightarrow+\infty$. Thus, because of the
last term of the right hand side in (4.4), we have that $\lim_{\mu
\rightarrow+\infty}S_{2n+1}^{\mu}=-\infty$. That is, negative entropies do
occur in the odd case. (Recall that $\left(  \mathbb{C},d\nu_{o,\mu}\right)  $
\textit{is not} a probability measure space for $\mu\ne0$.) Nevertheless we
will see now that for fixed $\mu>-\frac{1}{2}$ the sequence $\left\{
S_{2n+1}^{\mu}\right\}  _{n=0}^{\infty}$ is increasing and unbounded, and so
it is eventually positive. We have that%
\begin{align*}
S_{2n+3}^{\mu}  & =\left(  n+\frac{1}{2}\right)  \left(  \psi\left(
\mu+n+\frac{3}{2}\right)  +\psi\left(  n+1\right)  \right)  -\log\frac
{\gamma_{\mu}(2n+1)}{2^{2n+1}}\\
& +\left(  n+\frac{3}{2}\right)  \left(  \frac{1}{\mu+n+\frac{3}{2}}+\frac
{1}{n+1}\right)  +\psi\left(  \mu+n+\frac{3}{2}\right) \\
& +\psi\left(  n+1\right)  -\log\left(  \left(  n+1\right)  \left(
\mu+n+\frac{3}{2}\right)  \right) \\
& =S_{2n+1}^{\mu}+\left(  n+\frac{3}{2}\right)  \left(  \frac{1}{\mu
+n+\frac{3}{2}}+\frac{1}{n+1}\right) \\
& +\psi\left(  \mu+n+\frac{3}{2}\right)  -\log\left(  \mu+n+\frac{3}%
{2}\right)  +\psi\left(  n+1\right)  -\log\left(  n+1\right) \\
& >S_{2n+1}^{\mu}+\left(  n+\frac{3}{2}\right)  \left(  \frac{1}{\mu
+n+\frac{3}{2}}+\frac{1}{n+1}\right)  -\frac{1}{\mu+n+\frac{3}{2}}-\frac
{1}{n+1}\\
& =S_{2n+1}^{\mu}+\left(  n+\frac{1}{2}\right)  \left(  \frac{1}{\mu
+n+\frac{3}{2}}+\frac{1}{n+1}\right) \\
& >S_{2n+1}^{\mu},
\end{align*}
where we used Lemma 3.1(c). This proves that the sequence $\left\{
S_{2n+1}^{\mu}\right\}  _{n=0}^{\infty}$ is increasing. Moreover, since%
\begin{align*}
S_{2n+3}^{\mu}-S_{2n+1}^{\mu}  & =\left(  n+\frac{3}{2}\right)  \left(
\frac{1}{\mu+n+\frac{3}{2}}+\frac{1}{n+1}\right) \\
& +\psi\left(  \mu+n+\frac{3}{2}\right)  -\log\left(  \mu+n+\frac{3}{2}\right)
\\
& +\psi\left(  n+1\right)  -\log\left(  n+1\right)
\end{align*}
and by Lemma 3.1(c) we have that $\psi\left(  \mu+n+\frac{3}{2}\right)
-\log\left(  \mu+n+\frac{3}{2}\right)  \rightarrow0$ and also that
$\psi\left(  n+1\right)  -\log\left(  n+1\right)  \rightarrow0$ as
$n\rightarrow\infty$, then we conclude that \linebreak$\lim_{n\rightarrow
\infty}\left(  S_{2n+3}^{\mu}-S_{2n+1}^{\mu}\right)  =2$, which implies the
unboundedness of the sequence $\left\{  S_{2n+1}^{\mu}\right\}  _{n=0}%
^{\infty}$. This limit implies that $\lim_{n\rightarrow\infty}\frac
{S_{2n+1}^{\mu}}{2n+1}=1$, but a direct proof of this is as follows. Note
that
\begin{align*}
\frac{S_{2n+1}^{\mu}}{2n+1}  & =\frac{1}{2}\left(  \psi\left(  \mu+n+\frac
{3}{2}\right)  +\psi\left(  n+1\right)  \right)  -\log\frac{\left(
\gamma_{\mu}(2n+1)\right)  ^{\frac{1}{2n+1}}}{2n+1}\\
& -\log\left(  n+\frac{1}{2}\right) \\
& =\frac{1}{2}\left(  \psi\left(  \mu+n+\frac{3}{2}\right)  -\log\left(
n+\frac{1}{2}\right)  +\psi\left(  n+1\right)  -\log\left(  n+\frac{1}%
{2}\right)  \right) \\
& -\log\frac{\left(  \gamma_{\mu}(2n+1)\right)  ^{\frac{1}{2n+1}}}{2n+1}.
\end{align*}

Note that Lemmas 3.1(b) and 3.2 give us that $\psi\left(  \mu+n+\frac{3}%
{2}\right)  -\log\left(  n+\frac{1}{2}\right)  \rightarrow0$, $\psi\left(
n+1\right)  -\log\left(  n+\frac{1}{2}\right)  \rightarrow0$ and $\log
\frac{\left(  \gamma_{\mu}(2n+1)\right)  ^{\frac{1}{2n+1}}}{2n+1}%
\rightarrow-1$ as $n\rightarrow\infty$. Then we have that $\frac{S_{2n+1}%
^{\mu}}{2n+1}\rightarrow1$ as $n\rightarrow\infty$. \smallskip

Thus, we have proved the following theorem.

\bigskip

\textbf{Theorem 4.2 \ }\textit{The entropy of }$\xi_{2n+1}^{\mu}$\textit{\ is
given by}%
\[
S_{2n+1}^{\mu}=\left(  n+\frac{1}{2}\right)  \left(  \psi\left(  \mu
+n+\frac{3}{2}\right)  +\psi\left(  n+1\right)  \right)  -\log\frac
{\gamma_{\mu}(2n+1)}{2^{2n+1}},
\]
\textit{where }$\mu>-\frac{1}{2}$\textit{\ and }$n=0,1,2,...$\textit{. For
fixed }$\mu>-\frac{1}{2}$\textit{, the sequence }$\left\{  S_{2n+1}^{\mu
}\right\}  _{n=0}^{\infty}$\textit{\ is an unbounded increasing sequence such
that}%
\[
\lim_{n\rightarrow\infty}\left(  S_{2n+3}^{\mu}-S_{2n+1}^{\mu}\right)  =2,
\]
\textit{which implies that}
\[
\lim_{n\rightarrow\infty}\frac{S_{2n+1}^{\mu}}{2n+1}=1.
\]

\textit{For fixed }$n=0,1,2,...$\textit{, we have that}%
\[
\lim_{\mu\rightarrow+\infty}S_{2n+1}^{\mu}=-\infty.
\]

\bigskip

We can relate the entropies $S_{2n+1}^{\mu}$ with the entropies $S_{2n}^{\mu}$
as follows. We note that%
\begin{align*}
S_{2n+1}^{\mu}  & =\frac{2n+1}{2}\left(  \psi\left(  \mu+n+\frac{3}{2}\right)
+\psi\left(  n+1\right)  \right)  -\log\frac{\gamma_{\mu}(2n+1)}{2^{2n+1}}\\
& =\left(  n+\frac{1}{2}\right)  \left(  \frac{1}{\mu+n+\frac{1}{2}}%
+\psi\left(  \mu+n+\frac{1}{2}\right)  +\psi\left(  n+1\right)  \right) \\
& -\log\frac{\gamma_{\mu}(2n+1)}{2^{2n+1}}\\
& =\left(  1+\frac{1}{2n}\right)  \left(  S_{2n}^{\mu}+\log\frac{\gamma_{\mu
}(2n)}{2^{2n}}\right)  +\frac{n+\frac{1}{2}}{\mu+n+\frac{1}{2}}-\log
\frac{\gamma_{\mu}(2n+1)}{2^{2n+1}}\\
& =\left(  1+\frac{1}{2n}\right)  S_{2n}^{\mu}+\frac{n+\frac{1}{2}}%
{\mu+n+\frac{1}{2}}+\log\frac{\left(  \gamma_{\mu}(2n)\right)  ^{\frac{1}{2n}%
}}{2\left(  \mu+n+\frac{1}{2}\right)  }.
\end{align*}

So we have that%
\[
S_{2n+1}^{\mu}-S_{2n}^{\mu}=\frac{S_{2n}^{\mu}}{2n}+\frac{n+\frac{1}{2}}%
{\mu+n+\frac{1}{2}}+\log\frac{\left(  \gamma_{\mu}(2n)\right)  ^{\frac{1}{2n}%
}}{2\left(  \mu+n+\frac{1}{2}\right)  }.
\]

By using Lemma 3.2 and Theorem 4.1 we obtain%
\[
\lim_{n\rightarrow\infty}\left(  S_{2n+1}^{\mu}-S_{2n}^{\mu}\right)  =1.
\]

Similarly one has that%
\[
S_{2n}^{\mu}-S_{2n-1}^{\mu}=\frac{S_{2n-1}^{\mu}}{2n-1}+1+\log\frac
{\gamma_{\mu}(2n-1)^{\frac{1}{2n-1}}}{2n}.
\]

Lemma 3.2 and Theorem 4.2 allow us to conclude%
\[
\lim_{n\rightarrow\infty}\left(  S_{2n}^{\mu}-S_{2n-1}^{\mu}\right)  =1.
\]

Finally, observe that we can express the formulas (4.1) and (4.3) in terms of
the characteristic function $\theta$ of the odd positive integers as%
\[
S_{n}^{\mu}=\frac{n}{2}\left(  \psi\left(  \mu+\frac{n+\theta\left(  n\right)
+1}{2}\right)  +\psi\left(  \frac{n+\theta\left(  n+1\right)  +1}{2}\right)
\right)  -\log\frac{\gamma_{\mu}(n)}{2^{n}}.
\]

From this formula one can obtain at once the case $\mu=0$ (formula (1.1)) by
using the identity%
\[
\psi\left(  \frac{n+1}{2}\right)  +\psi\left(  \frac{n+2}{2}\right)
=2\psi\left(  n+1\right)  -2\log2,
\]
whose proof is an easy exercise by induction.

Combining Theorems 4.1 and 4.2 with the previous results, we have the following.

\bigskip

\textbf{Theorem 4.3 \ }\textit{Let }$\mu>-\frac{1}{2}$\textit{\ be fixed. The
entropy }$S_{n}^{\mu}$\textit{\ is given by}%
\[
S_{n}^{\mu}=\frac{n}{2}\left(  \psi\left(  \mu+\frac{n+\theta\left(  n\right)
+1}{2}\right)  +\psi\left(  \frac{n+\theta\left(  n+1\right)  +1}{2}\right)
\right)  -\log\frac{\gamma_{\mu}(n)}{2^{n}}.
\]
\textit{\ }

\textit{The sequence }$\left\{  S_{n}^{\mu}\right\}  _{n=0}^{\infty}%
$\textit{\ of entropies is such that the subsequences of even terms }$\left\{
S_{2n}^{\mu}\right\}  _{n=1}^{\infty}$\textit{\ and of odd terms }$\left\{
S_{2n+1}^{\mu}\right\}  _{n=0}^{\infty}$\textit{\ are increasing, the former
being positive and the latter being eventually positive. Moreover, we have
that}%
\[
\lim_{n\rightarrow\infty}\left(  S_{n+1}^{\mu}-S_{n}^{\mu}\right)  =1,
\]
\textit{which shows that the sequence }$\left\{  S_{n}^{\mu}\right\}
_{n=0}^{\infty}$\textit{\ is unbounded and implies that}%
\[
\lim_{n\rightarrow\infty}\frac{S_{n}^{\mu}}{n}=1.
\]

\bigskip

\section{Entropies in $L^{2}\left(  \mathbb{R},dg_{\mu}\right)  $}

\bigskip

Following the same sort of ideas we used in the previous section, we will
calculate in this section the entropies of monomials $t^{n}\in L^{2}\left(
\mathbb{R},dg_{\mu}\right)  $, $n=1,2,...$. (In the case $n=0$ we obtain from
the definition that $S_{L^{2}\left(  \mathbb{R},dg_{\mu}\right)  }\left(
1\right)  =0$.) That is, for $n=1,2,...$ we will calculate explicitly%
\begin{align*}
& S_{L^{2}\left(  \mathbb{R},dg_{\mu}\right)  }\left(  t^{n}\right) \\
& =\!\frac{1}{\Gamma\left(  \mu+\frac{1}{2}\right)  }\!\int_{\mathbb{R}%
}\!\left\vert t^{n}\right\vert ^{2}\log\left\vert t^{n}\right\vert ^{2}%
\!\exp\left(  -t^{2}\right)  \left\vert t\right\vert ^{2\mu}dt\!-\!\left\Vert
t^{n}\right\Vert _{L^{2}\left(  \mathbb{R},dg_{\mu}\right)  }^{2}%
\!\log\left\Vert t^{n}\right\Vert _{L^{2}\left(  \mathbb{R},dg_{\mu}\right)
}^{2}\\
& =\!\frac{1}{\Gamma\left(  \mu+\frac{1}{2}\right)  }\!\int_{0}^{\infty
}\!\!\!u^{n}\!\left(  \log u^{n}\right) \! \exp(-u)u^{\mu-\frac{1}{2}%
}du-\left\Vert t^{n}\right\Vert _{L^{2}\left(  \mathbb{R},dg_{\mu}\right)
}^{2}\log\left\Vert t^{n}\right\Vert _{L^{2}\left(  \mathbb{R},dg_{\mu
}\right)  }^{2}.
\end{align*}

A direct calculation gives us%
\[
\left\Vert t^{n}\right\Vert _{L^{2}\left(  \mathbb{R},dg_{\mu}\right)  }%
^{2}=\frac{\Gamma\left(  n+\mu+\frac{1}{2}\right)  }{\Gamma\left(  \mu
+\frac{1}{2}\right)  }.
\]

Next, define the function%
\[
\eta\left(  \alpha\right)  =\int_{0}^{\infty}u^{n\alpha}\exp\left(  -u\right)
u^{\mu-\frac{1}{2}}ds=\Gamma\left(  n\alpha+\mu+\frac{1}{2}\right)
\]
in a neighborhood of $\alpha=1$. By calculating the derivative $\eta^{\prime
}\left(  1\right)  $ in two different ways (as we did in previous section) we
find that%
\[
\int_{0}^{\infty}u^{n}\left(  \log u^{n}\right)  \exp\left(  -u\right)
u^{\mu-\frac{1}{2}}du=n\psi\left(  n+\mu+\frac{1}{2}\right)  \Gamma\left(
n+\mu+\frac{1}{2}\right)  .
\]

Then the entropy $S_{L^{2}\left(  \mathbb{R},dg_{\mu}\right)  }\left(
t^{n}\right)  $ is%
\begin{equation}
S_{L^{2}\left(  \mathbb{R},dg_{\mu}\right)  }\left(  t^{n}\right)
=\frac{\Gamma\left(  n+\mu+\frac{1}{2}\right)  }{\Gamma\left(  \mu+\frac{1}%
{2}\right)  }\left(  n\psi\left(  n+\mu+\frac{1}{2}\right)  -\log\frac
{\Gamma\left(  n+\mu+\frac{1}{2}\right)  }{\Gamma\left(  \mu+\frac{1}%
{2}\right)  }\right)  .\tag{5.1}%
\end{equation}

When $n=1$ formula (5.1) becomes%
\begin{align*}
S_{L^{2}\left(  \mathbb{R},dg_{\mu}\right)  }\left(  t\right)   &
=\frac{\Gamma\left(  \mu+\frac{3}{2}\right)  }{\Gamma\left(  \mu+\frac{1}%
{2}\right)  }\left(  \psi\left(  \mu+\frac{3}{2}\right)  -\log\frac
{\Gamma\left(  \mu+\frac{3}{2}\right)  }{\Gamma\left(  \mu+\frac{1}{2}\right)
}\right) \\
& =\left(  \mu+\frac{1}{2}\right)  \left(  \psi\left(  \mu+\frac{3}{2}\right)
-\log\left(  \mu+\frac{1}{2}\right)  \right)  .
\end{align*}

By using that $S_{L^{2}\left(  \mathbb{R},dg_{\mu}\right)  }\left(  t\right)
$ is homogeneous of degree $2$ we can calculate the entropy of the monomial
$\zeta_{1}^{\mu}\left(  t\right)  =\left(  \frac{2}{1+2\mu}\right)  ^{\frac
{1}{2}}t$, which is the second element of the canonical basis $\left\{
\zeta_{n}^{\mu}\right\}  _{n=0}^{\infty}$ of $L^{2}\left(  \mathbb{R},dg_{\mu
}\right)  $. In fact, we have that $S_{L^{2}\left(  \mathbb{R},dg_{\mu
}\right)  }\left(  \zeta_{1}^{\mu}\right)  =\frac{2}{1+2\mu}S_{L^{2}\left(
\mathbb{R},dg_{\mu}\right)  }\left(  t\right)  $, and then%
\begin{equation}
S_{L^{2}\left(  \mathbb{R},dg_{\mu}\right)  }\left(  \zeta_{1}^{\mu}\right)
=\psi\left(  \mu+\frac{3}{2}\right)  -\log\left(  \mu+\frac{1}{2}\right)
.\tag{5.2}%
\end{equation}

When $\mu=0$ this formula becomes%
\[
S_{L^{2}\left(  \mathbb{R},dg\right)  }\left(  \zeta_{1}^{0}\right)
=\psi\left(  \frac{3}{2}\right)  -\log\left(  \frac{1}{2}\right)
=2-\log2-\gamma,
\]
using $\psi\left(  \frac{3}{2}\right)  =2-2\log2-\gamma$, which is (1.2) as expected.

Unfortunately we can not continue the previous procedure in order to obtain
explicit formulas for the entropies of $\zeta_{n}^{\mu}\in L^{2}\left(
\mathbb{R},dg_{\mu}\right)  $ with $n\geq2$, since for those values of $n$ the
polynomials $\zeta_{n}^{\mu}$ are not longer monomials, and then (5.1) is not
useful. Nevertheless we will study some properties of the sequence $\left\{
s_{n}^{\mu}\right\}  _{n=0}^{\infty}$, where $s_{n}^{\mu}:=S_{L^{2}\left(
\mathbb{R},dg_{\mu}\right)  }\left(  t^{n}\right)  $, and compare them with
the results obtained in Section 4.

Before that, recall that the $\mu$-deformed Segal-Bargmann transform $B_{\mu
}:L^{2}\left(  \mathbb{R},dg_{\mu}\right)  \rightarrow\mathcal{B}_{\mu}^{2}$
is such that $B_{\mu}\left(  \zeta_{n}^{\mu}\right)  =\xi_{n}^{\mu}$,
$n=0,1,...$. When $n=0$ we have $\zeta_{0}^{\mu}\left(  t\right)  =1$,
$\xi_{0}^{\mu}\left(  z\right)  =1$, and $S_{L^{2}\left(  \mathbb{R},dg_{\mu
}\right)  }\left(  1\right)  =S_{L^{2}\left(  \mathbb{C},d\nu_{e,\mu}\right)
}\left(  1\right)  =0$. So in this case we see that $B_{\mu}$ preserves
entropy. Let us consider the case $n=1$. Formula (4.3) gives us%
\[
S_{L^{2}\left(  \mathbb{C},d\nu_{o,\mu}\right)  }\left(  B_{\mu}\left(
\zeta_{1}^{\mu}\right)  \right)  =\frac{1}{2}\left(  \psi\left(  \mu+\frac
{3}{2}\right)  +\psi\left(  1\right)  \right)  -\log\left(  \mu+\frac{1}%
{2}\right)  .
\]

This formula, $\psi\left(  1\right)  =-\gamma$, and (5.2) give us that%
\begin{equation}
S_{L^{2}\left(  \mathbb{C},d\nu_{o,\mu}\right)  }\left(  B_{\mu}\left(
\zeta_{1}^{\mu}\right)  \right)  -S_{L^{2}\left(  \mathbb{R},dg_{\mu}\right)
}\left(  \zeta_{1}^{\mu}\right)  =-\frac{1}{2}\left(  \psi\left(  \mu+\frac
{3}{2}\right)  +\gamma\right)  .\tag{5.3}%
\end{equation}

Observe that $\lim_{\mu\rightarrow-\frac{1}{2}^{+}}\left(  \psi\left(
\mu+\frac{3}{2}\right)  +\gamma\right)  =\psi\left(  1\right)  +\gamma=0$, and
that $\mu\mapsto\psi\left(  \mu+\frac{3}{2}\right)  +\gamma$ is an increasing
function since we have that $\psi^{\prime}\left(  x\right)  >0$ for $x>0$ (see
[M-O-S], p. 14). Then $\psi\left(  \mu+\frac{3}{2}\right)  +\gamma>0$ for
$\mu>-\frac{1}{2}$, and thus formula (5.3) tells us that $S_{L^{2}\left(
\mathbb{C},d\nu_{o,\mu}\right)  }\left(  B_{\mu}\left(  \zeta_{1}^{\mu
}\right)  \right)  <S_{L^{2}\left(  \mathbb{R},dg_{\mu}\right)  }\left(
\zeta_{1}^{\mu}\right)  $. That is, the $\mu$-deformed Segal-Bargmann
transform $B_{\mu}$ decreases the entropy of $\zeta_{1}^{\mu}$. We have
already noted that $B_{\mu}$ preserves the entropy of $\zeta_{0}^{\mu}$. It
seems reasonable to conjecture that $B_{\mu}$ increases the entropy of other
functions in $L^{2}\left(  \mathbb{R},dg_{\mu}\right)  $. (This is known to be
true in the case $\mu=0$. See [Snt1].)

As happens in the case of the sequence of entropies $\left\{  S_{n}^{\mu
}\right\}  _{n=0}^{\infty}$ in the previous section, the sequence of entropies
$\left\{  s_{n}^{\mu}\right\}  _{n=0}^{\infty}$ is unbounded as we will prove
now. By using the asymptotics%
\begin{equation}
\psi\left(  z\right)  =\log z+O\left(  z^{-1}\right)  ,\tag{5.4}%
\end{equation}
valid for $\left\vert \arg z\right\vert <\pi$ and $z\rightarrow\infty$ (see
[M-O-S], p. 18), and Stirling's formula%
\begin{equation}
\log\Gamma\left(  z\right)  =\left(  z-\frac{1}{2}\right)  \log z-z+O\left(
1\right)  ,\tag{5.5}%
\end{equation}
also valid for $\left\vert \arg z\right\vert <\pi$ and $z\rightarrow\infty$
(see [M-O-S], p. 12), we see that for large $n$ the term in parentheses in the
right hand side of (5.1) behaves like%
\begin{align*}
& n\left(  \log\left(  n+\mu+\frac{1}{2}\right)  +O\left(  n^{-1}\right)
\right)  -\left(  n+\mu\right)  \log\left(  n+\mu+\frac{1}{2}\right) \\
& +\left(  n+\mu+\frac{1}{2}\right)  +O(1)\\
& =n-\mu\log\left(  n+\mu+\frac{1}{2}\right)  +O\left(  1\right)  ,
\end{align*}
which is unbounded. In turn this implies that the sequence of entropies
$\left\{  s_{n}^{\mu}\right\}  _{n=0}^{\infty}$ is unbounded, as wanted.

Now let us see that the sequence $\left\{  s_{n}^{\mu}\right\}  _{n=0}%
^{\infty}$ is increasing (as the sequence $\left\{  S_{n}^{\mu}\right\}
_{n=0}^{\infty}$ is). First note that Lemma (3.1) (a) gives us%
\[
s_{1}^{\mu}=\left(  \mu+\frac{1}{2}\right)  \left(  \psi\left(  \mu+\frac
{3}{2}\right)  -\log\left(  \mu+\frac{1}{2}\right)  \right)  >0=s_{0}^{\mu},
\]
so let us prove that $s_{n+1}^{\mu}>s_{n}^{\mu}$ for $n\geq1$. Observe that%
\[
\frac{\Gamma\left(  n+\mu+\frac{1}{2}\right)  }{\Gamma\left(  \mu+\frac{1}%
{2}\right)  }=\prod_{k=1}^{n}\left(  k+\mu-\frac{1}{2}\right)  .
\]

So we can write (5.1) as%
\begin{align*}
s_{n}^{\mu}  & =\frac{\Gamma\left(  n+\mu+\frac{1}{2}\right)  }{\Gamma\left(
\mu+\frac{1}{2}\right)  }\left(  n\psi\left(  n+\mu+\frac{1}{2}\right)
-\log\prod_{k=1}^{n}\left(  k+\mu-\frac{1}{2}\right)  \right) \\
& =\frac{\Gamma\left(  n+\mu+\frac{1}{2}\right)  }{\Gamma\left(  \mu+\frac
{1}{2}\right)  }\sum_{k=1}^{n}\left(  \psi\left(  n+\mu+\frac{1}{2}\right)
-\log\left(  k+\mu-\frac{1}{2}\right)  \right)  .
\end{align*}

Then, by using Lemma (3.1) (a) we get%
\begin{align*}
s_{n+1}^{\mu}  & =\frac{\Gamma\left(  n+\mu+\frac{3}{2}\right)  }%
{\Gamma\left(  \mu+\frac{1}{2}\right)  }\sum_{k=1}^{n+1}\left(  \psi\left(
n+\mu+\frac{3}{2}\right)  -\log\left(  k+\mu-\frac{1}{2}\right)  \right) \\
& =\frac{\Gamma\left(  n+\mu+\frac{3}{2}\right)  }{\Gamma\left(  \mu+\frac
{1}{2}\right)  }\sum_{k=1}^{n}\left(
\begin{array}
[c]{c}%
\frac{1}{n+\mu+\frac{1}{2}}+\psi\left(  n+\mu+\frac{1}{2}\right) \\
-\log\left(  k+\mu-\frac{1}{2}\right)
\end{array}
\right) \\
& +\frac{\Gamma\left(  n+\mu+\frac{3}{2}\right)  }{\Gamma\left(  \mu+\frac
{1}{2}\right)  }\left(  \psi\left(  n+\mu+\frac{3}{2}\right)  -\log\left(
n+\mu+\frac{1}{2}\right)  \right) \\
& =\frac{n\Gamma\left(  n+\mu+\frac{1}{2}\right)  }{\Gamma\left(  \mu+\frac
{1}{2}\right)  }+\left(  n+\mu+\frac{1}{2}\right)  s_{n}^{\mu}\\
& +\frac{\Gamma\left(  n+\mu+\frac{3}{2}\right)  }{\Gamma\left(  \mu+\frac
{1}{2}\right)  }\left(  \psi\left(  n+\mu+\frac{3}{2}\right)  -\log\left(
n+\mu+\frac{1}{2}\right)  \right) \\
& >s_{n}^{\mu},
\end{align*}
which proves that $\left\{  s_{n}^{\mu}\right\}  _{n=0}^{\infty}$ is
increasing, as wanted. In particular we have that the sequence $\left\{
s_{n}^{\mu}\right\}  _{n=1}^{\infty}$ is positive. (Recall that since
$L^{2}\left(  \mathbb{R},dg_{\mu}\right)  $ is a probability measure space, we
have that $s_{n}^{\mu}\geq0$ for all $n=0,1,...$.)

Finally, observe that from formulas (2.1), (3.2) and (4.1) we see that (5.1)
can be written as%
\[
s_{n}^{\mu}=\frac{\Gamma\left(  n+\mu+\frac{1}{2}\right)  }{\Gamma\left(
\mu+\frac{1}{2}\right)  }\left(  S_{2n}^{\mu}-S_{n}\right)  .
\]

\bigskip

We know that $\lim_{n\rightarrow\infty}\frac{S_{2n}^{\mu}}{2n}=1$ (Theorem
4.1) and $\lim_{n\rightarrow\infty}\frac{S_{n}}{n}=1$ (Proposition 3.1). Then
we have that%
\[
\lim_{n\rightarrow+\infty}\frac{s_{n}^{\mu}}{n\Gamma\left(  n+\mu+\frac{1}%
{2}\right)  }=\left(  \Gamma\left(  \mu+\frac{1}{2}\right)  \right)  ^{-1}.
\]

So the sequence $\left\{  s_{n}^{\mu}\right\}  _{n=0}^{\infty}$ diverges to
infinity much faster than the sequence $\left\{  S_{n}^{\mu}\right\}
_{n=0}^{\infty}$ does.

\bigskip

\section{$\mu$-deformed Energies}

\bigskip

In this section we study two entropy-energy inequalities, known as
\textit{reverse log-Sobolev inequalities} (in the $\mu$-deformed
Segal-Bargmann space $\mathcal{B}_{\mu}^{2}$) that are proved in [A-S.1]. We
first quote the appropriate definition of energy from \linebreak\ [A-S.1] and
then calculate it for the functions in the canonical basis of $\mathcal{B}%
_{\mu}^{2}$. Since we already have calculated the entropies for these
functions, we can then proceed to the analysis of the two reverse log-Sobolev
inequalities.

\bigskip

\textbf{Definition 6.1 \ }\textit{For }$f\in\mathcal{B}_{e,\mu}^{2}%
$\textit{\ we define its }$\mu$\textit{-deformed energy }$E_{e,\mu}\left(
f\right)  $\textit{\ as}%
\[
E_{e,\mu}\left(  f\right)  =\int_{\mathbb{C}}\left\vert f\left(  z\right)
\right\vert ^{2}\left\vert z\right\vert ^{2}d\nu_{e,\mu}\left(  z\right)  .
\]

\textit{For }$f\in\mathcal{B}_{o,\mu}^{2}$\textit{\ we define its }$\mu
$\textit{-deformed energy }$E_{o,\mu}\left(  f\right)  $\textit{\ as}%
\[
E_{o,\mu}\left(  f\right)  =\int_{\mathbb{C}}\left\vert f\left(  z\right)
\right\vert ^{2}\left\vert z\right\vert ^{2}d\nu_{o,\mu}\left(  z\right)  .
\]

\textit{In general, for }$f\in\mathcal{B}_{\mu}^{2}$\textit{\ we define its
}$\mu$\textit{-deformed energy }$E_{\mu}\left(  f\right)  $\textit{\ as
}$E_{\mu}\left(  f\right)  =E_{e,\mu}\left(  f_{e}\right)  +E_{o,\mu}\left(
f_{o}\right)  $\textit{.}

\bigskip

We will denote by $E_{n}^{\mu}$ to the $\mu$-deformed energy $E_{\mu}\left(
\xi_{n}^{\mu}\right)  $, so we have $E_{2n}^{\mu}=E_{e,\mu}\left(  \xi
_{2n}^{\mu}\right)  $ and $E_{2n+1}^{\mu}=E_{o,\mu}\left(  \xi_{2n+1}^{\mu
}\right)  $. We have that%
\begin{align*}
E_{2n}^{\mu}  & =\frac{2^{\frac{1}{2}-\mu}}{\pi\Gamma\left(  \mu+\frac{1}%
{2}\right)  }\int_{\mathbb{C}}\left\vert \frac{z^{2n}}{\left(  \gamma_{\mu
}\left(  2n\right)  \right)  ^{\frac{1}{2}}}\right\vert ^{2}\left\vert
z\right\vert ^{2}K_{\mu-\frac{1}{2}}\left(  \left\vert z\right\vert
^{2}\right)  \left\vert z\right\vert ^{2\mu+1}dxdy\\
& =\frac{2^{\frac{1}{2}-\mu}2}{\Gamma\left(  \mu+\frac{1}{2}\right)
\gamma_{\mu}\left(  2n\right)  }\int_{0}^{\infty}K_{\mu-\frac{1}{2}}\left(
r^{2}\right)  r^{2(2n+\mu+2)}dr\\
& =\frac{2^{\frac{1}{2}-\mu}}{\Gamma\left(  \mu+\frac{1}{2}\right)
\gamma_{\mu}\left(  2n\right)  }\int_{0}^{\infty}K_{\mu-\frac{1}{2}}\left(
s\right)  s^{2n+\mu+\frac{3}{2}}ds.
\end{align*}

Since $2n+\mu+\frac{5}{2}>\left\vert \mu-\frac{1}{2}\right\vert $ we can use
formula (2.7) to write%
\[
E_{2n}^{\mu}=\frac{2^{\frac{1}{2}-\mu}}{\Gamma\left(  \mu+\frac{1}{2}\right)
\gamma_{\mu}\left(  2n\right)  }2^{2n+\mu+\frac{1}{2}}\Gamma\left(  n+\frac
{3}{2}\right)  \Gamma\left(  n+\mu+1\right)  ,
\]
which simplifies (by using (2.1)) to%
\begin{equation}
E_{2n}^{\mu}=\frac{2\Gamma\left(  n+\frac{3}{2}\right)  \Gamma\left(
n+\mu+1\right)  }{\Gamma\left(  n+1\right)  \Gamma\left(  n+\mu+\frac{1}%
{2}\right)  }.\tag{6.1}%
\end{equation}

Similarly we have that%
\begin{align*}
E_{2n+1}^{\mu}  & =\frac{2^{\frac{1}{2}-\mu}}{\pi\Gamma\left(  \mu+\frac{1}%
{2}\right)  }\int_{\mathbb{C}}\left\vert \frac{z^{2n+1}}{\left(  \gamma_{\mu
}\left(  2n+1\right)  \right)  ^{\frac{1}{2}}}\right\vert ^{2}\left\vert
z\right\vert ^{2}K_{\mu+\frac{1}{2}}\left(  \left\vert z\right\vert
^{2}\right)  \left\vert z\right\vert ^{2\mu+1}dxdy\\
& =\frac{2^{\frac{1}{2}-\mu}2}{\Gamma\left(  \mu+\frac{1}{2}\right)
\gamma_{\mu}\left(  2n+1\right)  }\int_{0}^{\infty}K_{\mu+\frac{1}{2}}\left(
r^{2}\right)  r^{2(2n+\mu+3)}dr\\
& =\frac{2^{\frac{1}{2}-\mu}}{\Gamma\left(  \mu+\frac{1}{2}\right)
\gamma_{\mu}\left(  2n+1\right)  }\int_{0}^{\infty}K_{\mu+\frac{1}{2}}\left(
s\right)  s^{2n+\mu+\frac{5}{2}}ds.
\end{align*}

Since $2n+\mu+\frac{7}{2}>\left\vert \mu+\frac{1}{2}\right\vert $ we can use
formula (2.7) to write%
\[
E_{2n+1}^{\mu}=\frac{2^{\frac{1}{2}-\mu}}{\Gamma\left(  \mu+\frac{1}%
{2}\right)  \gamma_{\mu}\left(  2n+1\right)  }2^{2n+\mu+\frac{3}{2}}%
\Gamma\left(  n+\frac{3}{2}\right)  \Gamma\left(  n+\mu+2\right)  ,
\]
which simplifies (by using (2.2)) to%
\begin{equation}
E_{2n+1}^{\mu}=\frac{2\Gamma\left(  n+\frac{3}{2}\right)  \Gamma\left(
n+\mu+2\right)  }{\Gamma\left(  n+1\right)  \Gamma\left(  n+\mu+\frac{3}%
{2}\right)  }.\tag{6.2}%
\end{equation}

When $\mu=0$ formulas (6.1) and (6.2) become%
\[
E_{2n}^{0}=2n+1\text{\qquad and\qquad}E_{2n+1}^{0}=2n+2,
\]
which agrees with the known result that the (undeformed) energy $E_{n}$ of the
function $\xi_{n}$ is $n+1$ (see [Snt1]).

In [A-S.1] the following two reverse log-Sobolev inequalities are proved in
the context of $\mu$-deformed Segal-Bargmann analysis (Theorems 5.1 and 5.2).

\bigskip

\textbf{Theorem 6.1 }\textit{For all }$c>1$\textit{\ there exists a real
number }$P_{e}\left(  c,\mu\right)  $\textit{\ such that for }$f\in
\mathcal{B}_{e,\mu}^{2}$\textit{\ we have}%
\[
E_{e,\mu}\left(  f\right)  \leq cS_{L^{2}\left(  \mathbb{C},d\nu_{e,\mu
}\right)  }\left(  f\right)  +P_{e}\left(  c,\mu\right)  \left\Vert
f\right\Vert _{\mathcal{B}_{e,\mu}^{2}}.
\]

\bigskip

\textbf{Theorem 6.2 }\textit{For all }$c>1$\textit{\ there exists a real
number }$P_{o}\left(  c,\mu\right)  $\textit{\ such that for }$f\in
\mathcal{B}_{o,\mu}^{2}$\textit{\ we have}%
\[
E_{o,\mu}\left(  f\right)  \leq cS_{L^{2}\left(  \mathbb{C},d\nu_{o,\mu
}\right)  }\left(  f\right)  +P_{o}\left(  c,\mu\right)  \left\Vert
f\right\Vert _{\mathcal{B}_{o,\mu}^{2}}.
\]

\bigskip

A direct consequence of these results is the following reverse log-Sobolev
inequality in the $\mu$-deformed Segal-Bargmann space $\mathcal{B}_{\mu}^{2}$
(Theorem 5.3 in \linebreak[4] [A-S.1]).

\bigskip

\textbf{Theorem 6.3 }\textit{For all }$c>1$\textit{\ there exists a real
number }$P\left(  c,\mu\right)  $\textit{\ such that for }$f\in\mathcal{B}%
_{\mu}^{2}$\textit{\ we have}%
\[
E_{\mu}\left(  f\right)  \leq c\left(  S_{L^{2}\left(  \mathbb{C},d\nu_{e,\mu
}\right)  }\left(  f_{e}\right)  +S_{L^{2}\left(  \mathbb{C},d\nu_{o,\mu
}\right)  }\left(  f_{o}\right)  \right)  +P\left(  c,\mu\right)  \left\Vert
f\right\Vert _{\mathcal{B}_{\mu}^{2}}.
\]

\bigskip

In particular, if we consider the elements $\xi_{n}^{\mu}$, $n=0,1,...$ of the
canonical basis of $\mathcal{B}_{\mu}^{2}$, Theorem 6.1 tells us that for all
$c>1$ there exists a constant $P_{e}\left(  c,\mu\right)  $ such that for all
$n=0,1,...$ we have that%
\begin{equation}
E_{2n}^{\mu}\leq cS_{2n}^{\mu}+P_{e}\left(  c,\mu\right)  ,\tag{6.3}%
\end{equation}
and Theorem 6.2 tells us that for all $c>1$ there exists a constant
$P_{o}\left(  c,\mu\right)  $ such that for all $n=0,1,...$ we have that%
\begin{equation}
E_{2n+1}^{\mu}\leq cS_{2n+1}^{\mu}+P_{o}\left(  c,\mu\right)  .\tag{6.4}%
\end{equation}

\textbf{Remark. }By using Stirling's formula it is easy to see from (6.1) and
(6.2) that for fixed $n=0,1,...$, we have that $E_{n}^{\mu}\rightarrow+\infty$
as $\mu\rightarrow+\infty$. We already know that $S_{2n+1}^{\mu}%
\rightarrow-\infty$ as $\mu\rightarrow+\infty$ (see Theorem 4.2). Then (6.4)
tells us that for any $c>1$ and any $n=0,1,...$ we have that $P_{o}\left(
c,\mu\right)  \geq E_{2n+1}^{\mu}-cS_{2n+1}^{\mu}\rightarrow+\infty$ as
$\mu\rightarrow+\infty$. That is, the values of the constant $P_{o}\left(
c,\mu\right)  $ in Theorem 6.2 will be as large as we want, by taking $\mu>0$
large enough.

So Theorem 6.1 tells us that $c>1$ is a sufficient condition to conclude the
existence of the constant $P_{e}\left(  c,\mu\right)  $ such that the
inequality (6.3) holds for all $n=0,1,...$. We will prove now that this
condition is also necessary, by showing that for fixed $\mu>-\frac{1}{2}$, the
sequence $\left\{  E_{2n}^{\mu}-cS_{2n}^{\mu}\right\}  _{n=0}^{\infty}$ is
bounded above if and only if $c>1$.

By using the formula%
\begin{equation}
\frac{\Gamma\left(  z+\alpha\right)  }{\Gamma\left(  z+\beta\right)
}=z^{\alpha-\beta}\left(  1+O\left(  z^{-1}\right)  \right)  ,\tag{6.5}%
\end{equation}
valid for $\left\vert \arg z\right\vert <\pi$ and $z\rightarrow\infty$ (see
[M-O-S], p. 12) we can write the following asymptotics for $E_{2n}^{\mu}$:%
\begin{align*}
E_{2n}^{\mu}  & =2\frac{\Gamma\left(  n+\frac{3}{2}\right)  }{\Gamma\left(
n+1\right)  }\frac{\Gamma\left(  n+\mu+1\right)  }{\Gamma\left(  n+\mu
+\frac{1}{2}\right)  }\\
& =2n^{\frac{1}{2}}\left(  1+O\left(  n^{-1}\right)  \right)  n^{\frac{1}{2}%
}\left(  1+O\left(  n^{-1}\right)  \right) \\
& =2n+O(1).
\end{align*}

Also, by using (2.1), (5.4), (5.5) and formula (4.1) for $S_{2n}^{\mu}$, we
can write%
\begin{align*}
S_{2n}^{\mu}  & =n\left(  \psi\left(  \mu+n+\frac{1}{2}\right)  +\psi\left(
n+1\right)  \right)  -\log\frac{\Gamma\left(  n+1\right)  \Gamma\left(
\mu+n+\frac{1}{2}\right)  }{\Gamma\left(  \mu+\frac{1}{2}\right)  }\\
& =n\left(  \log\left(  \mu+n+\frac{1}{2}\right)  +\log\left(  n+1\right)
+O\left(  n^{-1}\right)  \right) \\
& -\left(  n+\frac{1}{2}\right)  \log\left(  n+1\right)  +n+1\\
& -\left(  \mu+n\right)  \log\left(  \mu+n+\frac{1}{2}\right)  +\mu+n+\frac
{1}{2}+O\left(  1\right) \\
& =-\frac{1}{2}\log\left(  n+1\right)  -\mu\log\left(  \mu+n+\frac{1}%
{2}\right)  +2n+O\left(  1\right)  .
\end{align*}

Then we have that%

\begin{align*}
E_{2n}^{\mu}-cS_{2n}^{\mu}  & =2n+O(1)-c\left(
\begin{array}
[c]{c}%
-\frac{1}{2}\log\left(  n+1\right) \\
-\mu\log\left(  \mu+n+\frac{1}{2}\right) \\
+2n+O\left(  1\right)
\end{array}
\right) \\
& =\left(  1-c\right)  2n+\frac{c}{2}\log\left(  n+1\right)  +c\mu\log\left(
\mu+n+\frac{1}{2}\right)  +O\left(  1\right) \\
& =\left(  1-c\right)  2n+\frac{c}{2}\log\frac{n+1}{\mu+n+\frac{1}{2}}\\
& +c\left(  \mu+\frac{1}{2}\right)  \log\left(  \mu+n+\frac{1}{2}\right)
+O\left(  1\right) \\
& =\left(  1-c\right)  2n+c\left(  \mu+\frac{1}{2}\right)  \log\left(
\mu+n+\frac{1}{2}\right)  +O\left(  1\right)  .
\end{align*}

Clearly, the sequence $\left\{  E_{2n}^{\mu}-cS_{2n}^{\mu}\right\}
_{n=0}^{\infty}$ is bounded above if and only if $c>1$. This shows that the
condition $c>1$ is the best possible in the reverse log-Sobolev inequality in
Theorem 6.1, namely that this inequality does not hold for $c\leq1$.

Now let us consider Theorem 6.2. We know that $c>1$ is a sufficient condition
to conclude the existence of the constant $P_{o}\left(  c,\mu\right)  $ such
that the inequality (6.4) holds for all $n=0,1,...$. We will see now that this
condition is also necessary, by showing that for fixed $\mu>-\frac{1}{2}$, the
sequence $\left\{  E_{2n+1}^{\mu}-cS_{2n+1}^{\mu}\right\}  _{n=0}^{\infty}$ is
bounded above if and only if $c>1$.

By using (6.5) we can write the following asymptotics for $E_{2n+1}^{\mu}$:%
\begin{align*}
E_{2n+1}^{\mu}  & =\frac{2\Gamma\left(  n+\frac{3}{2}\right)  \Gamma\left(
n+\mu+2\right)  }{\Gamma\left(  n+1\right)  \Gamma\left(  n+\mu+\frac{3}%
{2}\right)  }\\
& =2n^{\frac{1}{2}}\left(  1+O\left(  n^{-1}\right)  \right)  n^{\frac{1}{2}%
}\left(  1+O\left(  n^{-1}\right)  \right) \\
& =2n+O(1).
\end{align*}

By using (2.2), (5.4), (5.5) and formula (4.3) for $S_{2n+1}^{\mu}$, we can write%
\begin{align*}
S_{2n+1}^{\mu}  & =\left(  n+\frac{1}{2}\right)  \left(  \psi\left(
\mu+n+\frac{3}{2}\right)  +\psi\left(  n+1\right)  \right) \\
& -\log\frac{\Gamma\left(  n+1\right)  \Gamma\left(  \mu+n+\frac{3}{2}\right)
}{\Gamma\left(  \mu+\frac{1}{2}\right)  }\\
& =\left(  n+\frac{1}{2}\right)  \left(  \log\left(  \mu+n+\frac{3}{2}\right)
+\log\left(  n+1\right)  +O\left(  n^{-1}\right)  \right) \\
& -\left(  n+\frac{1}{2}\right)  \log\left(  n+1\right)  +n+1\\
& -\left(  \mu+n+1\right)  \log\left(  \mu+n+\frac{3}{2}\right)  +\mu
+n+\frac{3}{2}+O(1)\\
& =-\left(  \mu+\frac{1}{2}\right)  \log\left(  \mu+n+\frac{3}{2}\right)
+2n+O\left(  1\right)  .
\end{align*}

Then we have that%
\begin{align*}
E_{2n+1}^{\mu}-cS_{2n+1}^{\mu}  & =2n+O(1)-c\left(  \!\!%
\begin{array}
[c]{c}%
-\left(  \mu+\frac{1}{2}\right)  \log\left(  \mu+n+\frac{3}{2}\right) \\
+2n+O\left(  1\right)
\end{array}
\!\!\right) \\
& =\left(  1-c\right)  2n+c\left(  \mu+\frac{1}{2}\right)  \log\left(
\mu+n+\frac{3}{2}\right)  +O\left(  1\right)  .
\end{align*}

As in the case of Theorem 6.1 considered above, we see now that the sequence
$\left\{  E_{2n+1}^{\mu}-cS_{2n+1}^{\mu}\right\}  _{n=0}^{\infty}$ is bounded
above if and only if $c>1$, which shows that the condition $c>1$ is the best
possible in the reverse log-Sobolev inequality in Theorem 6.2.

Either one of the two cases considered in this section shows that the
condition $c>1$ in Theorem 6.3 is also the best possible.

\bigskip

\section{Final remarks}

\bigskip

In conclusion, we have just a few comments.

Firstly, it would be interesting to evaluate in closed form the entropies of
the elements of the canonical basis of $L^{2}\left(  \mathbb{R},dg_{\mu
}\right)  $. This has not even been done yet in the case $\mu=0$.

Secondly, we would like to repeat the conjecture that the $\mu$-deformed
Segal-Bargmann transform increases the entropy of some functions. And again,
this is plausible since it is known to be true when $\mu=0$. (See [Snt1].)

\bigskip

\section*{Acknowledgments}

\bigskip

The first author wishes to thank the Universidad Panamericana (Mexico City)
for giving him the support for being a full-time doctoral student at CIMAT
from 2001 to 2004. Both authors thank Shirley Bromberg, Pavel Naumkin, Roberto
Quezada Batalla and Carlos Villegas-Blas for valuable comments.

\bigskip


\end{document}